\documentclass[a4paper]{article}

\usepackage[pages=all, color=black, position={current page.south}, placement=bottom, scale=1, opacity=1, vshift=5mm]{background}\SetBgContents{}
\usepackage[margin=1in]{geometry} % full-width

\RequirePackage{fix-cm}

\usepackage{amsmath,amsfonts,amssymb,epsfig}
\usepackage{graphicx}
\usepackage{amssymb}
\usepackage{mathptmx}      % use Times fonts if available on your TeX system
\usepackage{mathtools} 
\usepackage{algorithm}
\usepackage[noend]{algpseudocode}
\usepackage{faktor}

\usepackage[T1]{fontenc}
\usepackage[utf8]{inputenc}
\usepackage[english]{babel}
\usepackage{setspace}
\usepackage{amsthm}
\usepackage{array}
\usepackage{float}
\usepackage{graphicx}
\usepackage{bm}
\usepackage{fancyhdr}
\usepackage{longtable}
\usepackage{epstopdf}
\usepackage{hyperref}
\usepackage{tikz}
\usepackage{pgfplots}
\usepackage{color}
\usepackage{float}
\usepackage{multicol}
\usepackage{subcaption}
\usepackage{latexsym}
\usepackage{hyperref}
\usepackage{cancel}

\usepackage{listings}

\lstdefinestyle{custompython}{
  belowcaptionskip=1\baselineskip,
  breaklines=true,
  frame=L,
  xleftmargin=\parindent,
  language=python,
  showstringspaces=false,
  basicstyle=\footnotesize\ttfamily,
  keywordstyle=\bfseries\color{green!40!black},
  commentstyle=\itshape\color{purple!40!black},
  identifierstyle=\color{blue},
  stringstyle=\color{orange},
}
\title{Physics Informed Neural Network Code for 2D Transient Problems (PINN-2DT) Compatible with Google Colab}
\author{Pawe\l{} Maczuga$^1$, Maciej Sikora$^1$,
Maciej Skocze\'n$^1$, \\ Przemys\l{}aw Ro\.znawski$^1$, Filip T\l{}uszcz$^1$, Marcin Szubert$^1$, \\
Marcin \L{}o\'s$^1$, Witold Dzwinel$^1$,
Keshav Pingali$^2$, Maciej Paszy\'nski$^1$}

\date{$^1$AGH University of Krak\'ow, Poland \\
$^2$ Oden Institute, The University of Texas at Austin, USA}

\begin{document}

\maketitle

\begin{abstract}
We present an open-source Physics Informed Neural Network environment for simulations of transient phenomena on two-dimensional rectangular domains, with the following features: (1) it is compatible with Google Colab which allows automatic execution on cloud environment; (2) it supports two dimensional time-dependent PDEs; (3) it provides simple interface for definition of the residual loss, boundary condition and initial loss, together with their weights; (4) it support Neumann and Dirichlet boundary conditions; (5) it allows for customizing the number of layers and neurons per layer, as well as for arbitrary activation function; (6) the learning rate and number of epochs are available as parameters; (7) it automatically differentiates PINN with respect to spatial and temporal variables; (8) it provides routines for plotting the convergence (with running average), initial conditions learnt, 2D and 3D snapshots from the simulation and movies (9) it includes a library of problems: (a) non-stationary heat transfer; (b) wave equation modeling a tsunami;
(c) atmospheric simulations including thermal inversion;
(d) tumor growth simulations.
\end{abstract}


\section{Program summary}

\emph{Program Title:} PINN-2DT

%Program Files doi: http://dx.doi.org/10.17632/pbpsyzyvfy.1

\emph{Licensing provisions:} MIT license (MIT)

\emph{Programming language:} Python

\emph{Nature of problem:} Solving non-stationary problems in 2D

\emph{Solution method:} Physics Informed Neural Networks. The implementation requires definition of PDE loss, initial conditions loss, and boundary conditions loss

\emph{Additional comments including Restrictions and Unusual features:} The code is prepared in a way to be compatible with Google Colab 
\section{Introduction}

The goal of this paper is to replace the functionality of the time-dependent solver we published using isogeometric analysis and fast alternating directions solver \cite{IGAADS,IGAADS2,IGAADS3} with the Physics Informed Neural Network (PINN) python library that can be easily executed on Colab.
The PINN proposed in 2019 by Prof. Karniadakis revolutionized the way in which neural networks find solutions to initial-value problems described using partial differential equations \cite{c1}
This method treats the neural network as a function approximating the solution of the given partial differential equation $u (x) = PINN (x)$. After computing the necessary differential operators, the neural network and its appropriate differential operators are inserted into the partial differential equation. The residuum of the partial differential equation and the boundary-initial conditions are assumed as the loss function. The learning process involves sampling the loss function at different points by calculating the PDE residuum, the residuum of the initial conditions, as well as the residuum of the boundary conditions.
The PINN methodology has had exponential growth in the number of papers and citations since its creation in 2019.
It has multiple applications, from solid mechanics \cite{Solid}, geology  \cite{Groundwater}, medical applications \cite{Tumor}, and even the 
phase-field modeling of fracture \cite{Fracture}.
%Why use PINN solvers instead of classical or higher order finite element methods (e.g., isogeometric analysis) solvers?
PINN/VPINN solvers have affordable computational costs. 
They can be easily implemented using pre-existing libraries and environments (like Pytorch and Google Colab).
They are easily parallelizable, especially on GPU. They have great approximation capabilities, and they enable finding solutions to a family of problems.
There are several modern stochastic optimizers  \cite{ADAMS,Optimization,ADAMW,Lions}, where the ADAM algorithm seems to be the most popular \cite{ADAMS}.
They easily find high-quality minimizers of the loss functions employed.

In this paper, we present the PINN library with the following features
\begin{itemize}
\item It is implemented in Pythorch and compatible with Google Colab.
\item It supports two-dimensional problems defined on a rectangular domain.
\item It is suitable for smooth problems without singularities resulting from large contrast material data.
\item It enables the definition of the PDE residual loss function in the space-time domain.
\item It supports the loss function for defining the initial condition.
\item It provides loss functions for Neumann and Dirichlet boundary conditions.
\item It allows for customization of the loss functions and their weights.
\item It allows for defining an arbitrary number of layers of the neural network and an arbitrary number of neurons per layer.
\item The learning rate, the kind of activation function, and a number of epochs are problem-specific parameters.
\item It automatically performs differentiation of the PINN with respect to spatial and temporal variables.
\item It provides tools for plotting the convergence of all the loss functions, together with the running average.
\item It enables the plotting of the exact and learned initial conditions.
\item It plots 2D or 3D  snapshots from the simulations.
\item It generates gifs with the simulation animation.
\end{itemize}
We illustrate our PINN-2DT code with four numerical examples. The first one concerns the model heat transfer problem. The second one presents the solution to the wave equation. The third one is the simulation of the thermal inversion, and the last one is the simulation of brain tumor growth.

There are the following available PINN libraries.
First and most important is the DeepXDE library \cite{PINNlib0} by the team of Prof. Karniadakis. It is an extensive library with huge functionality, including ODEs, PDEs, complex geometries, different initial and boundary conditions, and forward and inverse problems. 
It supports several tensor libraries such as TensorFlow, PyTorch, JAX, and PaddlePaddle.

Another interesting library is 
IDRLnet \cite{PINNlib1}. It uses pytorch, numpy, and Matplotlib.
This library is illustrated on four different examples, namely the wave equation, Allan-Cahn equations, Volterra integrodifferential equations, and variational minimization problems. 

What is the novelty of our library?
Our library is very simple to use and compatible with Google Colab.
It is a natural ``copy" of the functionality of the IGA-ADS library \cite{IGAADS} into the PINN methodology.
It contains a simple, straightforward interface for solving different time-dependent problems. Our library can be executed without accessing the HPC center just by using the Colab functionality.

The structure of the paper is the following.
In Section 2, we recall the general idea of PINN on the example of the heat transfer problem. Section 3 is devoted to our code structure, from Colab implementation, model parameters, basic Python classes, how we define initial and boundary conditions, loss functions, how we run the training, and how we process the output. Section 4 provides four examples from heat transfer, wave equation, thermal inversion, and tumor growth simulations. We conclude the paper in Section 5.

\section{Physics Informed Neural Network for transient problems on the example of heat transfer problem}

Let us consider a strong form of the exemplary transient PDE, the heat transfer problem.
Find $u\in C^2(0,1)$ for $(x,y) \in \Omega = [0,1]^2$, $t\in [0,T]$ such that
\begin{eqnarray}
\underbrace{\frac{\partial u(x,y,t)}{\partial t}}_{\textrm{temperature evolution}}
\underbrace{-\epsilon \frac{\partial ^2u(x,y,t)}{\partial x^2}-\epsilon \frac{\partial ^2u(x,y,t)}{\partial y^2}}_{\textrm{diffusion term}}=\underbrace{f(x,y,t)}_{\textrm{forcing}}, (x,y,t) \in \Omega\times[0,T],
\end{eqnarray}
with initial condition \begin{eqnarray}
u(x,y,0)=u_0(x,y)
\end{eqnarray}
and zero-Neumann boundary condition
\begin{eqnarray}
\frac{\partial u}{\partial n}=0 \textrm{ } (x,y)\in \partial \Omega
\end{eqnarray}

In the Physics Informed Neural Network approach, the neural network is the solution, namely 
\begin{eqnarray}
u(x,y,t)=PINN(x,y,t)=A_n \sigma\left(A_{n-1}\sigma(...\sigma(A_1\begin{bmatrix}x \\ y \\ t \end{bmatrix}+B_1)...+B_{n-1}\right)+B_n
\end{eqnarray}
where $A_i$ are matrices representing DNN layers, $B_i$ represent bias vectors, and $\sigma$ is the non-linear activation function, e.g.,  sigmoid, which as we have shown in \cite{Activation}, is the best choice for PINN.
We define the loss function as the residual of the PDE

\begin{eqnarray}
LOSS_{PDE}(x,y,t) = 
\left(\frac{\partial PINN(x,y,t)}{\partial t}
-\epsilon \frac{\partial ^2 PINN(x,y,t)}{\partial x^2}-\epsilon \frac{\partial^2 PINN(x,y,t)}{\partial y^2}-f(x,y,t)\right)^2
\end{eqnarray}

We also define the loss for training the initial condition as the residual of the initial condition

\begin{eqnarray}
LOSS_{Init}(x,y,0) = 
\left(PINN(x,y,0)-u_0(x,y)\right)^2
\end{eqnarray}
as well as the loss of the residual of the boundary condition

\begin{eqnarray}
LOSS_{BC}(x,y,t) = 
\left(\frac{\partial PINN(x,y,t)}{\partial n}-0\right)^2
\end{eqnarray}

The sketch of the training procedure is the following

\begin{itemize}
\item Repeat
\begin{itemize}
\item Select points $(x,y,t) \in \Omega \times [0,T]$ randomly
\item Correct the weights using the strong loss
\begin{eqnarray}
A^k_{i,j} = A^k_{i,j} - \eta \frac{\partial LOSS_{PDE}(x,y,t)}{\partial A^k_{i,j}} \\
B^k_{i} = B^k_{i} - \eta \frac{\partial
LOSS_{PDE}(x,y,t)}{\partial B^k_{i}} 
\end{eqnarray}
where $\eta \in (0,1)$ is the training rate. 
\item Select point $(x,y) \in \partial \Omega$ randomly
\begin{eqnarray}
A^k_{i,j} = A^k_{i,j} - \eta \frac{\partial LOSS_{BC}(x,y,t)}{\partial A^k_{i,j}} \\
B^k_{i} = B^k_{i} - \eta \frac{\partial
LOSS_{BC}(x,y,t)}{\partial B^k_{i}} 
\end{eqnarray}
where $\eta \in (0,1)$. 
\item Select point $(x,y,0) \in \Omega \times \{ 0 \}$ randomly
\begin{eqnarray}
A^k_{i,j} = A^k_{i,j} - \eta \frac{\partial LOSS_{Init}(x,y,0)}{\partial A^k_{i,j}} \\
B^k_{i} = B^k_{i} - \eta \frac{\partial
LOSS_{Init}(x,y,0)}{\partial B^k_{i}} 
\end{eqnarray}
where $\eta \in (0,1)$. 
\end{itemize}
\item Until $w_{PDE} LOSS_{PDE}+w_{BC}LOSS_{BC}+w_{Init}LOSS_{Init} \leq \delta$
\end{itemize}
In practice, this simple stochastic gradient method is replaced by a more sophisticated e.g., ADAM method \cite{ADAMS}.

\section{Structure of the code}

\subsection{Colab implementation}

Our code is available at

{\tt https://github.com/pmaczuga/pinn-notebooks}

The code can be downloaded, openned in Google Colab, and executed in the fully automatic mode.

The code has been created to be compatible with Google Colab, and it employs the pytorch library.

\begin{lstlisting}
from typing import Callable
import matplotlib.pyplot as plt
import numpy as np
import torch
...
\end{lstlisting}

The code can automatically run on a cluster of GPUs, as provided by the Google Colab computing environment

\begin{lstlisting}
device = torch.device("cuda" if torch.cuda.is_available() else "cpu")
\end{lstlisting}

\subsection{Parameters}

There are the following model parameters that the user can define
\begin{itemize}
\item {\tt LENGTH}, {\tt TOTAL\_TIME}. The code works in the space-time domain, where the training is performed by selecting point along $x$, $y$
and $t$ axes. The {\tt LENGTH} parameter defines the dimension of the domain along $x$ and $y$ axes. The domain dimension is {\tt [0,LENGTH]x[0,LENGTH]x[0,TOTAL\_TIME]}.
The {\tt TOTAL\_TIME} parameter defines the length of the space-time domain along the $t$ axis. It is the total time of the transient phenomena we want to simulate.
\item {\tt N\_POINTS}. This parameter defines the number of points used for training. By default, the points are selected randomly along $x$, $y$, and $t$ axes. It is easily possible to extend the code to support different numbers of points or different distributions of points along different axes of the coordinate system.
\item {\tt N\_POINTS\_PLOT}. This parameter defines the number of points used for probing the solution and plotting the output plots after the training.
\item {\tt WEIGHT\_RESIDUAL}, {\tt WEIGHT\_INITIAL}, {\tt WEIGHT\_BOUNDARY}. These parameters define the weights for the training of residual, initial condition, and boundary condition loss functions.
\item {\tt LAYERS}, {\tt NEURONS\_PER\_LAYER}. These parameters define the neural network by providing the number of layers and number of neurons per neural network layer.
\item {\tt EPOCHS}, and {\tt LEARNING\_RATE} provide a number of epochs and the training rate for the training procedure.
\end{itemize}

Below we provide the exemplary values of the parameters as employed for the wave equation simulations
\begin{lstlisting}
## Parameters
LENGTH = 2. 
TOTAL_TIME = .5 
N_POINTS = 15 
N_POINTS_PLOT = 150 WEIGHT_RESIDUAL = 0.03 
WEIGHT_INITIAL = 1.0 
WEIGHT_BOUNDARY = 0.0005 
LAYERS = 10
NEURONS_PER_LAYER = 120
EPOCHS = 150_000
LEARNING_RATE = 0.00015
GRAVITY=9.81
\end{lstlisting}

\subsection{PINN class}

The PINN class defines the functionality for a simple neural network accepting three features as input, namely the values of $(x,y,t)$ and returning a single output, namely the value of the solution $u(x,y,t)$.
We provide the following features:
\begin{itemize}
\item The {\tt f} routine compute the values of the approximate solution at point $(x,y,t)$.
\item 
The routines {\tt dfdt}, {\tt dfdx}, {\tt dfdy} compute the derivatives of the approximate solution at point $(x,y,t)$ with respect to either $x$, $y$, or $t$ using the pytorch autograd method.
\end{itemize}

\begin{lstlisting}
class PINN(nn.Module):
    def __init__(self, num_hidden: int, dim_hidden: int, act=nn.Tanh()):

def f(pinn: PINN, x: torch.Tensor, y: torch.Tensor, t: torch.Tensor) -> torch.Tensor:
    return pinn(x, y, t)


def df(output: torch.Tensor, input: torch.Tensor, order: int = 1) -> torch.Tensor:
    df_value = output
    for _ in range(order):
        df_value = torch.autograd.grad(
            df_value,
            input,
            grad_outputs=torch.ones_like(input),
            create_graph=True,
            retain_graph=True,
        )[0]

    return df_value


def dfdt(pinn: PINN, x: torch.Tensor, y: torch.Tensor, t: torch.Tensor, order: int = 1):
    f_value = f(pinn, x, y, t)
    return df(f_value, t, order=order)


def dfdx(pinn: PINN, x: torch.Tensor, y: torch.Tensor, t: torch.Tensor, order: int = 1):
    f_value = f(pinn, x, y, t)
    return df(f_value, x, order=order)

def dfdy(pinn: PINN, x: torch.Tensor, y: torch.Tensor, t: torch.Tensor, order: int = 1):
    f_value = f(pinn, x, y, t)
    return df(f_value, y, order=order)
\end{lstlisting}

\subsection{Processing initial and boundary conditions}

Since the training is performed in the space-time domain 
{\tt [0,LENGTH]x[0,LENGTH]x[0,TOTAL\_TIME]}, we provide in 
\begin{itemize}
    \item 
{\tt get\_interior\_points} the functionality to identify the points from the training of the residual loss, in
\item {\tt get\_initial\_points} the functionality to identify points for the training of the initial loss, and in 
\item {\tt get\_boundary\_points} the functionality for training the boundary loss.
\end{itemize}

\begin{lstlisting}
def get_boundary_points(x_domain, y_domain, t_domain, n_points, \
      device = torch.device("cpu"), requires_grad=True):
    """
         .+------+
       .' |    .'|
      +---+--+'  |
      |   |  |   |
    y |  ,+--+---+
      |.'    | .' t
      +------+'
         x
    """
    x_linspace = torch.linspace(x_domain[0], x_domain[1], n_points)
    y_linspace = torch.linspace(y_domain[0], y_domain[1], n_points)
    t_linspace = torch.linspace(t_domain[0], t_domain[1], n_points)

    x_grid, t_grid = torch.meshgrid( x_linspace, t_linspace, indexing="ij")
    y_grid, _      = torch.meshgrid( y_linspace, t_linspace, indexing="ij")

    x_grid = x_grid.reshape(-1, 1).to(device)
    x_grid.requires_grad = requires_grad
    y_grid = y_grid.reshape(-1, 1).to(device)
    y_grid.requires_grad = requires_grad
    t_grid = t_grid.reshape(-1, 1).to(device)
    t_grid.requires_grad = requires_grad

    x0 = torch.full_like(t_grid, x_domain[0], requires_grad=requires_grad)
    x1 = torch.full_like(t_grid, x_domain[1], requires_grad=requires_grad)
    y0 = torch.full_like(t_grid, y_domain[0], requires_grad=requires_grad)
    y1 = torch.full_like(t_grid, y_domain[1], requires_grad=requires_grad)

    down    = (x_grid, y0,     t_grid)
    up      = (x_grid, y1,     t_grid)
    left    = (x0,     y_grid, t_grid)
    right   = (x1,     y_grid, t_grid)

    return down, up, left, right
\end{lstlisting}

\begin{lstlisting}
def get_initial_points(x_domain, y_domain, t_domain, n_points, \
      device = torch.device("cpu"), requires_grad=True):
    x_linspace = torch.linspace(x_domain[0], x_domain[1], n_points)
    y_linspace = torch.linspace(y_domain[0], y_domain[1], n_points)
    x_grid, y_grid = torch.meshgrid( x_linspace, y_linspace, indexing="ij")
    x_grid = x_grid.reshape(-1, 1).to(device)
    x_grid.requires_grad = requires_grad
    y_grid = y_grid.reshape(-1, 1).to(device)
    y_grid.requires_grad = requires_grad
    t0 = torch.full_like(x_grid, t_domain[0], requires_grad=requires_grad)
    return (x_grid, y_grid, t0)
\end{lstlisting}

\begin{lstlisting}
def get_interior_points(x_domain, y_domain, t_domain, n_points, \
      device = torch.device("cpu"), requires_grad=True):
    x_raw = torch.linspace(x_domain[0], x_domain[1], steps=n_points, requires_grad=requires_grad)
    y_raw = torch.linspace(y_domain[0], y_domain[1], steps=n_points, requires_grad=requires_grad)
    t_raw = torch.linspace(t_domain[0], t_domain[1], steps=n_points, requires_grad=requires_grad)
    grids = torch.meshgrid(x_raw, y_raw, t_raw, indexing="ij")

    x = grids[0].reshape(-1, 1).to(device)
    y = grids[1].reshape(-1, 1).to(device)
    t = grids[2].reshape(-1, 1).to(device)

    return x, y, t
\end{lstlisting}

\subsection{Loss functions}

Inside the {\tt Loss} class, we provide interfaces for the definition of the loss functions.
Namely, we define the {\tt residual\_loss}, {\tt initial\_loss} and {\tt boundary\_loss}. Since the initial and boundary loss is universal, and residual loss is problem specific, we provide fixed implementations for the initial and boundary losses, assuming that the initial state is prescribed in the {\tt initial\_condition} routine and that the boundary conditions are zero Neumann. The code can be easily extended to support different boundary conditions.

\begin{lstlisting}
class Loss:

...

    def residual_loss(self, pinn: PINN):
        x, y, t = get_interior_points(self.x_domain, self.y_domain, self.t_domain, \
             self.n_points, pinn.device())
        u = f(pinn, x, y, t)
        z = self.floor(x, y)
        loss = #HERE DEFINE RESIDUAL LOSS
        return loss.pow(2).mean()

    def initial_loss(self, pinn: PINN):
        x, y, t = get_initial_points(self.x_domain, self.y_domain, self.t_domain, \ 
             self.n_points, pinn.device())
        pinn_init = self.initial_condition(x, y)
        loss = f(pinn, x, y, t) - pinn_init
        return loss.pow(2).mean()

    def boundary_loss(self, pinn: PINN):
        down, up, left, right = get_boundary_points(self.x_domain, self.y_domain, self.t_domain,  \
             self.n_points, pinn.device())
        x_down,  y_down,  t_down    = down
        x_up,    y_up,    t_up      = up
        x_left,  y_left,  t_left    = left
        x_right, y_right, t_right   = right

        loss_down  = dfdy( pinn, x_down,  y_down,  t_down  )
        loss_up    = dfdy( pinn, x_up,    y_up,    t_up    )
        loss_left  = dfdx( pinn, x_left,  y_left,  t_left  )
        loss_right = dfdx( pinn, x_right, y_right, t_right )

        return loss_down.pow(2).mean()  + \
            loss_up.pow(2).mean()    + \
            loss_left.pow(2).mean()  + \
            loss_right.pow(2).mean()
\end{lstlisting}

The initial condition is defined in the {\tt initial\_condition} routine, which returns a value of the initial condition at point $(x,y,0)$.

\begin{lstlisting}
# Initial condition
def initial_condition(x: torch.Tensor, y: torch.Tensor) -> torch.Tensor:
...
    res = #HERE DEFINE THE INITIAL CONDITION AT (x,y,0)
    return res
\end{lstlisting}

\subsection{Training}

During the training, we select the Adam \cite{ADAMS} optimizer, and we prescribe that for every 1000 epochs of training, we will write the summary of the values of the residual, initial, and boundary losses. The user can modify this optimizer and the reporting frequency.

\begin{lstlisting}
def train_model(
    nn_approximator: PINN,
    loss_fn: Callable,
    learning_rate: int = 0.01,
    max_epochs: int = 1_000
) -> PINN:

    optimizer = torch.optim.Adam(nn_approximator.parameters(), lr=learning_rate)
    loss_values = []
    residual_loss_values = []
    initial_loss_values = []
    boundary_loss_values = []

    start_time = time.time()

    for epoch in range(max_epochs):

        try:

            loss: torch.Tensor = loss_fn(nn_approximator)
            optimizer.zero_grad()
            loss[0].backward()
            optimizer.step()

            loss_values.append(loss[0].item())
            residual_loss_values.append(loss[1].item())
            initial_loss_values.append(loss[2].item())
            boundary_loss_values.append(loss[3].item())
            if (epoch + 1) % 1000 == 0:
                epoch_time = time.time() - start_time
                start_time = time.time()

                print(f"Epoch: {epoch + 1} - Loss: {float(loss[0].item()):>7f}, \
                     Residual Loss: {float(loss[1].item()):>7f}, \
                     Initital Loss: {float(loss[2].item()):>7f}, \
                     Boundary Loss: {float(loss[3].item()):>7f}")

        except KeyboardInterrupt:
            break

    return nn_approximator, np.array(loss_values), \
       np.array(residual_loss_values), \
       np.array(initial_loss_values), \
       np.array(boundary_loss_values)
\end{lstlisting}

\subsection{Output}

\begin{figure}
      \centering
      \includegraphics[width=0.4\textwidth]{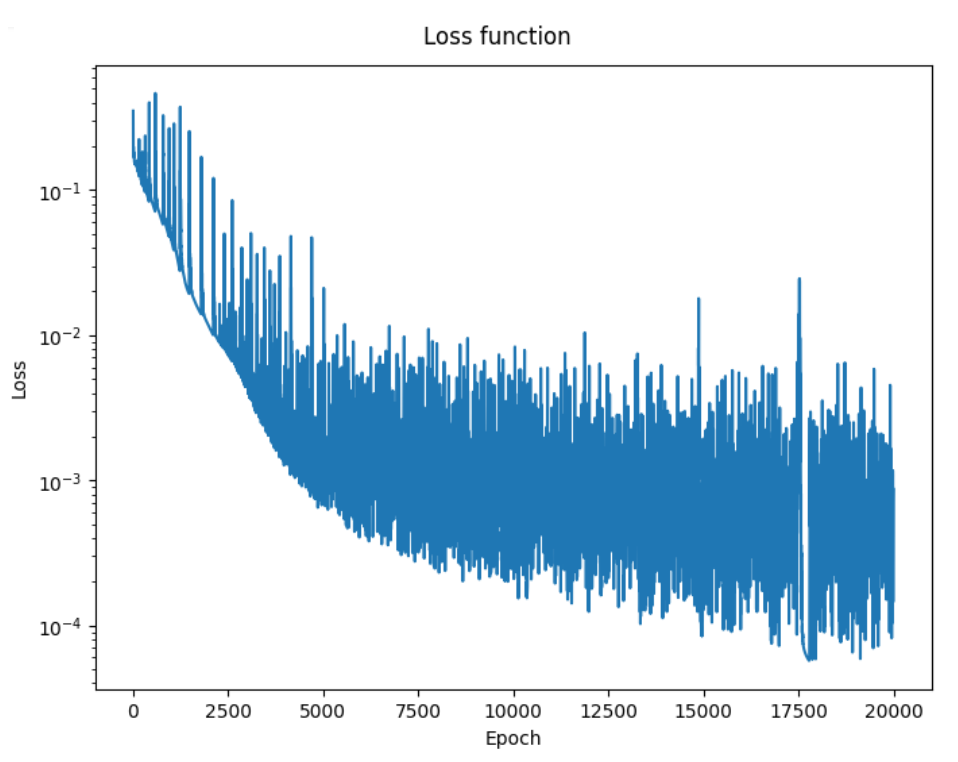}
      \caption{Heat equation. Convergence of the residual loss function.}
      \label{fig:ex1}
\end{figure}

We provide several routines for plotting the convergence of the loss function (see Fig. \ref{fig:ex1}, 

\begin{lstlisting}
# Plotting

# Loss function
average_loss = running_average(loss_values, window=100)
fig, ax = plt.subplots(figsize=(8, 6), dpi=100)
ax.set_title("Loss function (runnig average)")
ax.set_xlabel("Epoch")
ax.set_ylabel("Loss")
ax.plot(average_loss)
ax.set_yscale('log')
\end{lstlisting}

\begin{figure}
      \centering
      \includegraphics[width=0.4\textwidth]{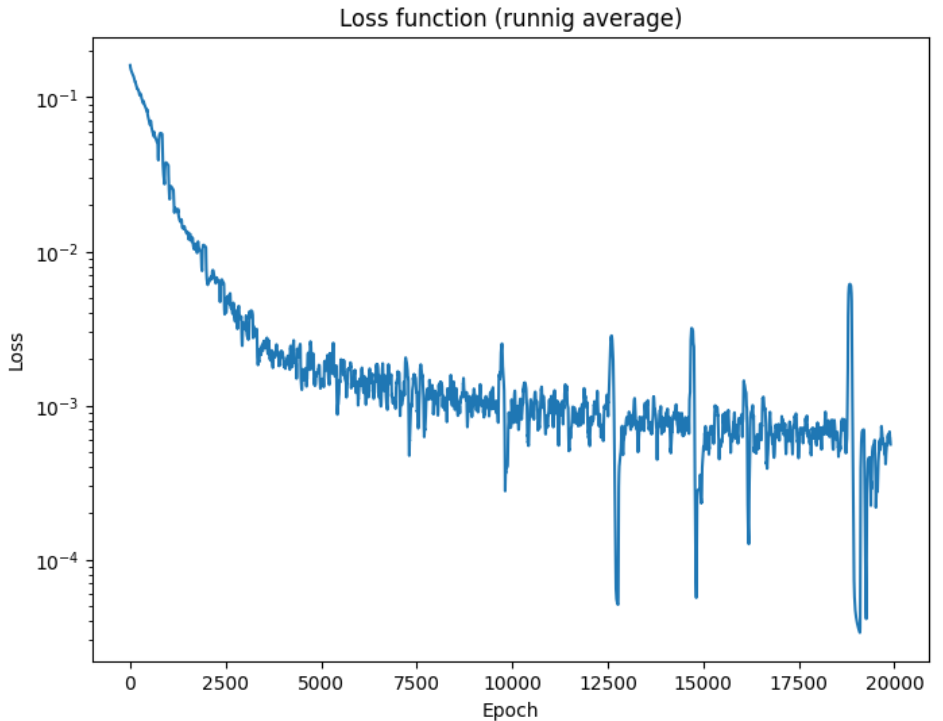}
      \caption{Heat equation. Running average from the convergence of the residual loss function.}
      \label{fig:ex2}
\end{figure}

\begin{figure}
      \centering
      \includegraphics[width=0.8\textwidth]{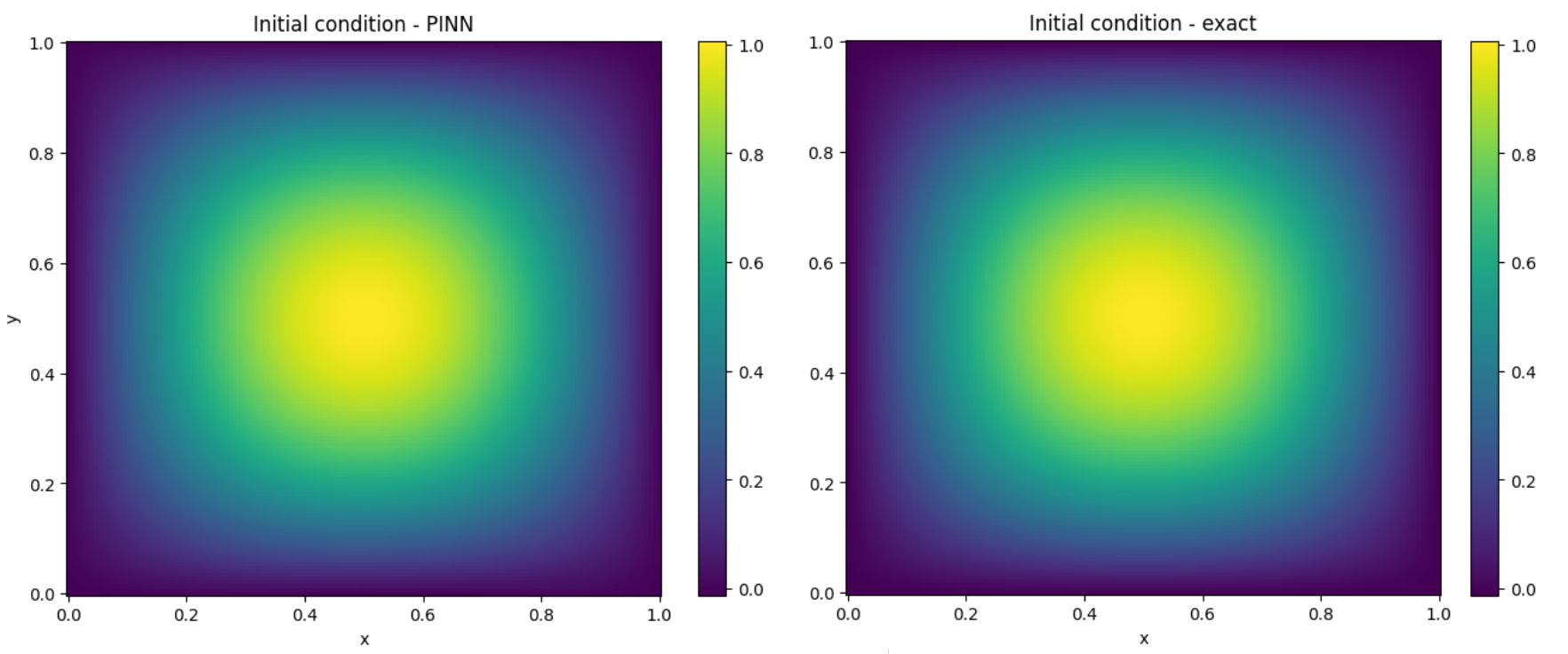}
      \caption{Heat equation. Initial conditions in 2D.}
      \label{fig:ex3}
\end{figure}

for plotting the running average of the loss (see Fig. \ref{fig:ex2}),  

\begin{lstlisting}
average_loss = running_average(initial_loss_values, window=100)
fig, ax = plt.subplots(figsize=(8, 6), dpi=100)
ax.set_title("Initial loss function (running average)")
ax.set_xlabel("Epoch")
ax.set_ylabel("Loss")
ax.plot(average_loss)
ax.set_yscale('log')
\end{lstlisting}

for plotting the initial conditions in 2D (see Fig. \ref{fig:ex3})

\begin{lstlisting}
# Plotting
# Initial condition
base_dir = '.'
x, y, _ = get_initial_points(x_domain, y_domain, t_domain, N_POINTS_PLOT, requires_grad=False)
z = initial_condition(x, y)
fig = plot_color(z, x, y, N_POINTS_PLOT, N_POINTS_PLOT, "Initial condition - exact")
t_value = 0.0
t = torch.full_like(x, t_value)
z = pinn(x, y, t)
fig = plot_color(z, x, y, N_POINTS_PLOT, N_POINTS_PLOT, "Initial condition - PINN")
\end{lstlisting}

\begin{figure}
      \centering
      \includegraphics[width=0.8\textwidth]{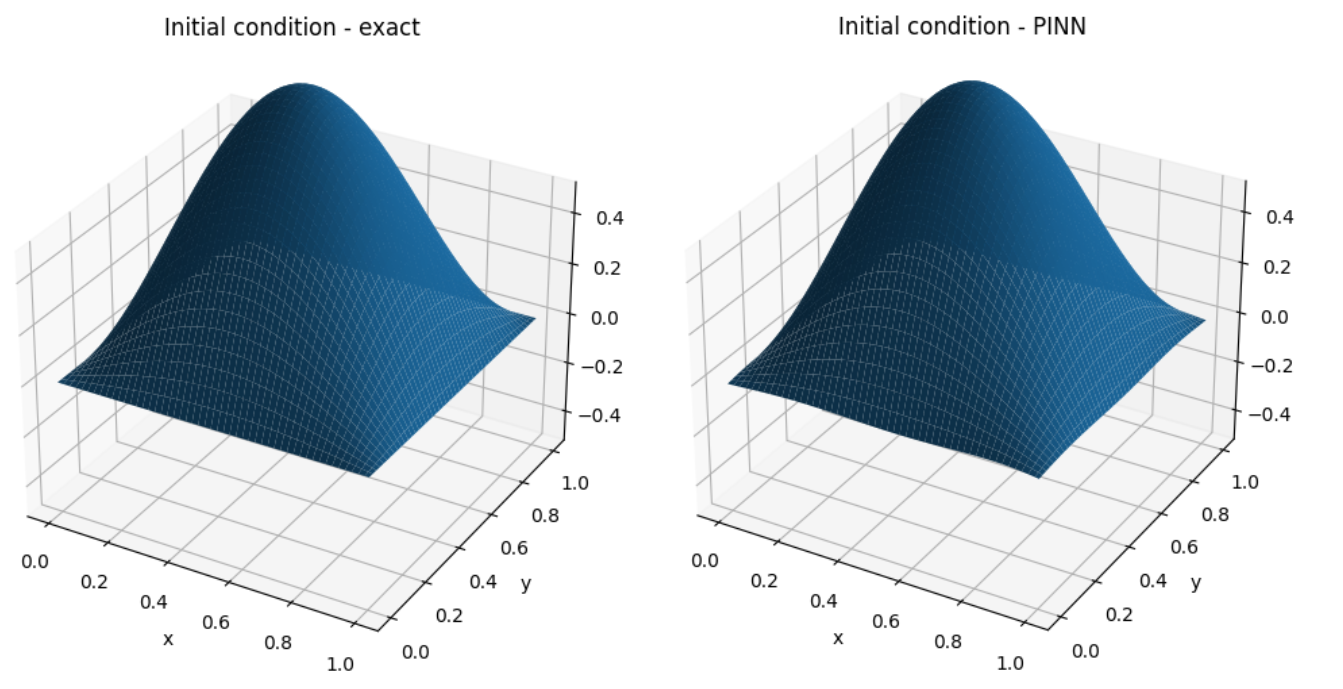}
      \caption{Heat equation. Initial conditions in 3D.}
      \label{fig:ex4}
\end{figure}

for plotting the initial conditions in 3D (see Fig. \ref{fig:ex4})

\begin{lstlisting}
# Plotting
# Initial condition
x, y, _ = get_initial_points(x_domain, y_domain, t_domain, N_POINTS_PLOT, requires_grad=False)
z = initial_condition(x, y)
fig = plot_3D(z, x, y, N_POINTS_PLOT, N_POINTS_PLOT, "Initial condition - exact")
z = pinn(x, y ,t)
fig = plot_3D(z, x, y, N_POINTS_PLOT, N_POINTS_PLOT, f"Initial condition - pinn")

\end{lstlisting}

\begin{figure}
      \centering
      \includegraphics[width=0.5\textwidth]{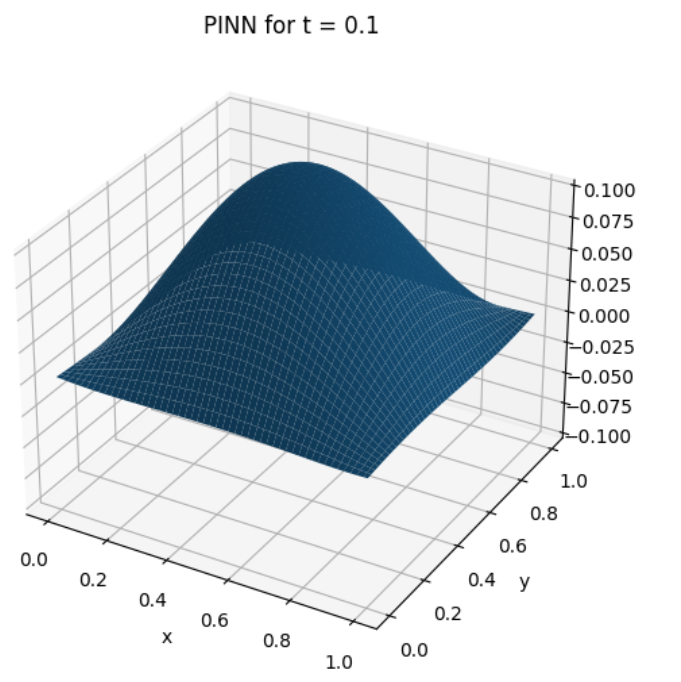}
      \caption{Heat equation. Snapshot from the simulation.}
      \label{fig:ex5}
\end{figure}

for plotting the snapshots of the solution (see Fig. \ref{fig:ex5})

\begin{lstlisting}
def plot(idx, t_value):
    t = torch.full_like(x, t_value)
    z = pinn(x, y, t)
    fig = plot_color(z, x, y, N_POINTS_PLOT, N_POINTS_PLOT, f"PINN for t = {t_value}")
    fig = plot_3D(z, x, y, N_POINTS_PLOT, N_POINTS_PLOT, f"PINN for t = {t_value}")
    #plt.savefig(base_dir + '/img/img_{:03d}.png'.format(idx))

time_values = np.arange(0, TOTAL_TIME, 0.01)
for idx, t_val in enumerate(time_values):
    plot(idx, t_value)
    z = pinn(x, y, t)
    fig = plot_color(z, x, y, N_POINTS_PLOT, N_POINTS_PLOT, f"PINN for t = {t_value}")
    fig = plot_3D(z, x, y, N_POINTS_PLOT, N_POINTS_PLOT, f"PINN for t = {t_val}")
    #plt.savefig(base_dir + '/img/img_{:03d}.png'.format(idx))

time_values = np.arange(0, TOTAL_TIME, 0.01)

for idx, t_val in enumerate(time_values):
    plot(idx, t_val)
\end{lstlisting}

and for the generation of the animated gif with the simulation results.

\begin{lstlisting}
from google.colab import drive
drive.mount('/content/drive')
import imageio
frames = []
for idx in range(len(time_values)):
    image = imageio.v2.imread(base_dir + '/img/img_{:03d}.png'.format(idx))
    frames.append(image)

imageio.mimsave('./tsunami_wave12.gif', # output gif
                frames, # array of input frames
                duration=0.1) # optional: frames per secon
\end{lstlisting}

\section{Examples of the instantiation}

\subsection{Heat transfer}

In this section, we present the numerical results for the model heat transfer problem described in Section 2.
The residual loss function
$LOSS_{PDE}(x,y,t) = 
\left(\frac{\partial PINN(x,y,t)}{\partial t}
-\frac{\partial ^2 PINN(x,y,t)}{\partial x^2}- \frac{\partial^2 PINN(x,y,t)}{\partial y^2}-f(x,y,t)\right)^2$
translates into the following code
\begin{lstlisting}
    def residual_loss(self, pinn: PINN):
        x, y, t = get_interior_points(self.x_domain, 
        self.y_domain, self.t_domain,        self.n_points, pinn.device())
        u = f(pinn, x, y, t)
        z = self.floor(x, y)
        loss = dfdt(pinn, x, y, t, order=1) - \
               dfdx(pinn, x, y, t, order=2)  - \
               dfdy(pinn, x, y, t, order=2)  
\end{lstlisting}

We employ the manufactured solution technique, where we assume the solution of the following form
\begin{eqnarray}
    u(x,y,t)=\exp^{-2\Pi^2 t}\sin{\Pi x}\sin{\Pi y}
\end{eqnarray}
over $\Omega = [0,1]^2$
To obtain this particular solution, we set up the zero Dirichlet boundary conditions, which require the following code

\begin{lstlisting}
def boundary_loss_dirichlet(self, pinn: PINN):
        down, up, left, right = get_boundary_points(self.x_domain, 
        self.y_domain, self.t_domain, self.n_points, pinn.device())
        x_down,  y_down,  t_down    = down
        x_up,    y_up,    t_up      = up
        x_left,  y_left,  t_left    = left
        x_right, y_right, t_right   = right
        loss_down  = f( pinn, x_down,  y_down,  t_down  )
        loss_up    = f( pinn, x_up,    y_up,    t_up    )
        loss_left  = f( pinn, x_left,  y_left,  t_left  )
        loss_right = f( pinn, x_right, y_right, t_right )
        return loss_down.pow(2).mean()  + \
            loss_up.pow(2).mean()    + \
            loss_left.pow(2).mean()  + \
            loss_right.pow(2).mean()
\end{lstlisting}

We also setup the initial state
\begin{eqnarray}
    u_0(x,y)=\sin{(\Pi x)}\sin{(\Pi y)}
\end{eqnarray}
which translates into the following code
\begin{lstlisting}
def initial_condition(x: torch.Tensor, y: torch.Tensor) -> torch.Tensor:
    res = torch.sin(torch.pi*x) * torch.sin(torch.pi*y)   
    return res
\end{lstlisting}

\begin{figure}
      \centering
      \includegraphics[width=0.6\textwidth]{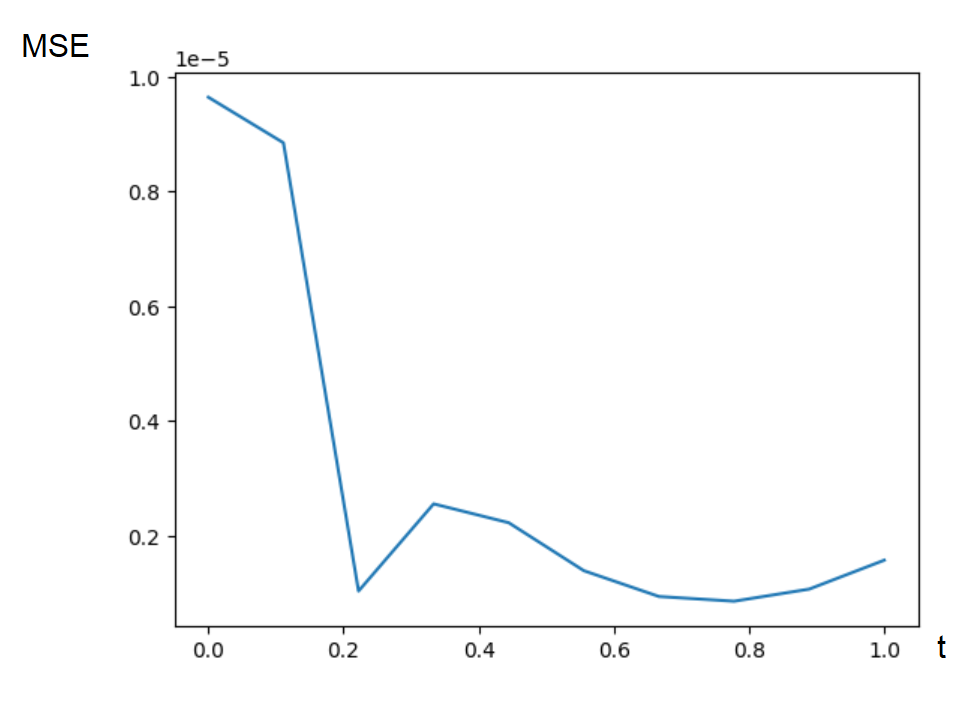}
      \caption{Heat equation. Numerical error of the trained PINN solution to the heat transfer problem with manufactured solution.}
      \label{fig:MSE}
\end{figure}

The default setup of the parameters for this simulation is the following:

\begin{lstlisting}
LENGTH = 1.
TOTAL_TIME = 1.
N_POINTS = 15
N_POINTS_PLOT = 150
WEIGHT_RESIDUAL = 1.0
WEIGHT_INITIAL = 1.0
WEIGHT_BOUNDARY = 1.0
LAYERS = 4
NEURONS_PER_LAYER = 80
EPOCHS = 20_000
LEARNING_RATE = 0.002
\end{lstlisting}

The convergence of the loss function is presented in Fig. \ref{fig:ex1}. The running average of the loss is presented in Fig. \ref{fig:ex2}. The comparison of exact and trained initial conditions is presented in Fig. \ref{fig:ex3} in 2D and Fig. \ref{fig:ex4} in 3D. The snapshot from the simulation is presented in Fig. \ref{fig:ex5} for time moment $t=0.1$.
The mean square error of the computed simulation is presented in Fig. \ref{fig:MSE}. We can see the high accuracy of the trained PINN results.

\subsection{Wave equations}

In this section, we consider the simulations of the wave equation, as motivated by the finite element method model described in \cite{Tsunami}.
Let us consider a strong form of the wave equation:
Find $u\in C^2(0,1)$ for $(x,y) \in \Omega = [0,1]^2$, $t\in [0,T]$ such that
\begin{eqnarray}
\frac{\partial^2 u(x,y,t)}{\partial t^2}-
\left(\frac{\partial }{\partial x},\frac{\partial }{\partial y}\right) \cdot \left(g (u(x,y,t)-z(x,y))\left(\frac{\partial u}{\partial x},\frac{\partial u}{\partial y}\right)\right)=0, \quad (x,y,t) \in \Omega\times[0,T],
\end{eqnarray}
with initial condition \begin{eqnarray}
u(x,y,0)=u_0(x,y), \textrm{  } (x,y)\in \Omega
\end{eqnarray}
and zero-Neumann boundary condition
\begin{eqnarray}
\frac{\partial u(x,y,t)}{\partial n}=0, \textrm{  } (x,y,t)\in \partial \Omega \times [0,T]
\end{eqnarray}
Here $z(x,y)$ stands for the seabed and seashore topography, $g=9.81$ is the acceleration due to the Earth gravity, and $u_0(x,y)$ is the initial shape of the wave.
We expand the wave equation by computing the partial derivatives. 

In our simulation, we run the wave propagation in the ``swimming pool"; thus, we assume $z(x,y)=0$. It implies some simplifications in the PDE

\begin{eqnarray}
\frac{\partial^2 u(x,y,t)}{\partial t^2}-
\left( g
\left(\frac{\partial u(x,y,t)}{\partial x}-\cancel{\frac{\partial z(x,y)}{\partial x}}\right)
\frac{\partial u(x,y,t)}{\partial x}
\right)
-\left( g
\left(u(x,y,t)-z(x,y)\right)
\frac{\partial^2 u(x,y,t)}{\partial x^2}
\right) \notag \\
-
\left(g \left( \frac{\partial u(x,y,t)}{\partial y} -\cancel{\frac{\partial z(x,y)}{\partial y}} \right)\frac{\partial u(x,y,t)}{\partial y}\right)
-
\left(g \left(  u(x,y,t)-z(x,y) \right)\frac{\partial^2 u(x,y,t)}{\partial y^2}\right)=0
\end{eqnarray}

In the Physics Informed Neural Network approach, the neural network represents the solution,  
\begin{eqnarray}
u(x,y,t)=PINN(x,y,t)=A_n \sigma\left(A_{n-1}\sigma(...\sigma(A_1 \begin{bmatrix} x \\ y \\ t\end{bmatrix}+B_1)...+B_{n-1}\right)+B_n
\end{eqnarray}
with $A_i$ being the matrices representing layers, $B_i$ are vectors representing biases, and $\sigma$ is sigmoid activation function \cite{Activation}.
We define the loss function as the residual of the PDE

\begin{eqnarray}
LOSS_{PDE}(x,y,t) = 
\left(
\frac{\partial^2 PINN(x,y,t)}{\partial t^2}-
 g
\left(\frac{\partial PINN(x,y,t)}{\partial x}\right)^2
-g
\left(PINN(x,y,t)-z(x,y)\right)
\frac{\partial^2 PINN(x,y,t)}{\partial x^2}
\right. \notag \\
\left. 
-g 
\left(\frac{\partial PINN(x,y,t)}{\partial y}\right)^2
-
g \left(  PINN(x,y,t)-z(x,y) \right)
\frac{\partial^2 PINN(x,y,t)}{\partial y^2}
\right)^2
\end{eqnarray}

This residual translates into the following code

\begin{lstlisting}
def residual_loss(self, pinn: PINN):
  x, y, t = get_interior_points(self.x_domain, self.y_domain, self.t_domain, 
  self.n_points, pinn.device())
  u = f(pinn, x, y, t)
  z = self.floor(x, y)
  loss = dfdt(pinn, x, y, t, order=2) - GRAVITY * ( dfdx(pinn, x, y, t) ** 2
  + (u-z) * dfdx(pinn, x, y, t, order=2) + dfdy(pinn, x, y, t) ** 2   
  + (u-z) * dfdy(pinn, x, y, t, order=2)
  )
  return loss.pow(2).mean() 
\end{lstlisting}

We also define the loss for training of the initial condition. It is defined as the residual of the initial condition

\begin{eqnarray}
LOSS_{Init}(x,y,0) = 
\left(PINN(x,y,0)-u_0(x,y)\right)^2
\end{eqnarray}

Similarly, we define the loss of the residual of the boundary conditions

\begin{eqnarray}
LOSS_{BC}(x,y,t) = 
\left(\frac{\partial PINN(x,y,t)}{\partial n}(x,y,t)-0\right)^2
\end{eqnarray}

We do not have to change the code for the initial and boundary conditions, we just provide an implementation of the initial state

\begin{lstlisting}
def initial_condition(x: torch.Tensor, y: torch.Tensor) -> torch.Tensor:
  r = torch.sqrt((x-LENGTH/2)**2 + (y-LENGTH/2)**2)
  res = 2 * torch.exp(-(r)**2 * 30) + 2
  return res
\end{lstlisting}

\begin{figure}
      \centering
      \includegraphics[width=0.5\textwidth]{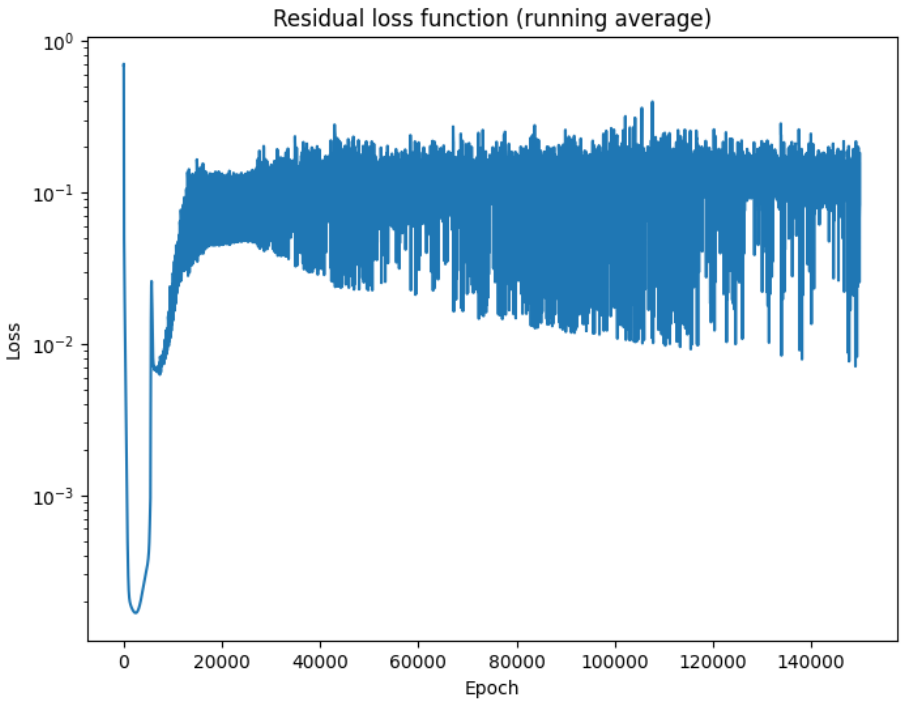}
      \caption{Wave equation. Convergence of the loss function.}
      \label{fig:losses}
\end{figure}

\begin{figure}
      \centering
      \includegraphics[width=0.4\textwidth]{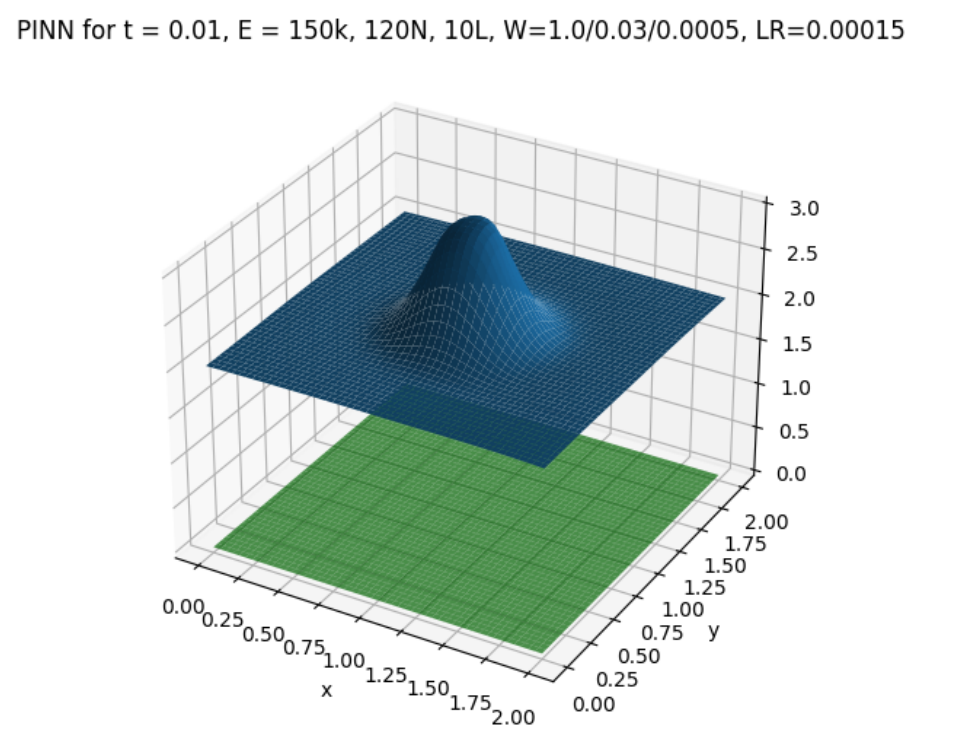}    \includegraphics[width=0.4\textwidth]{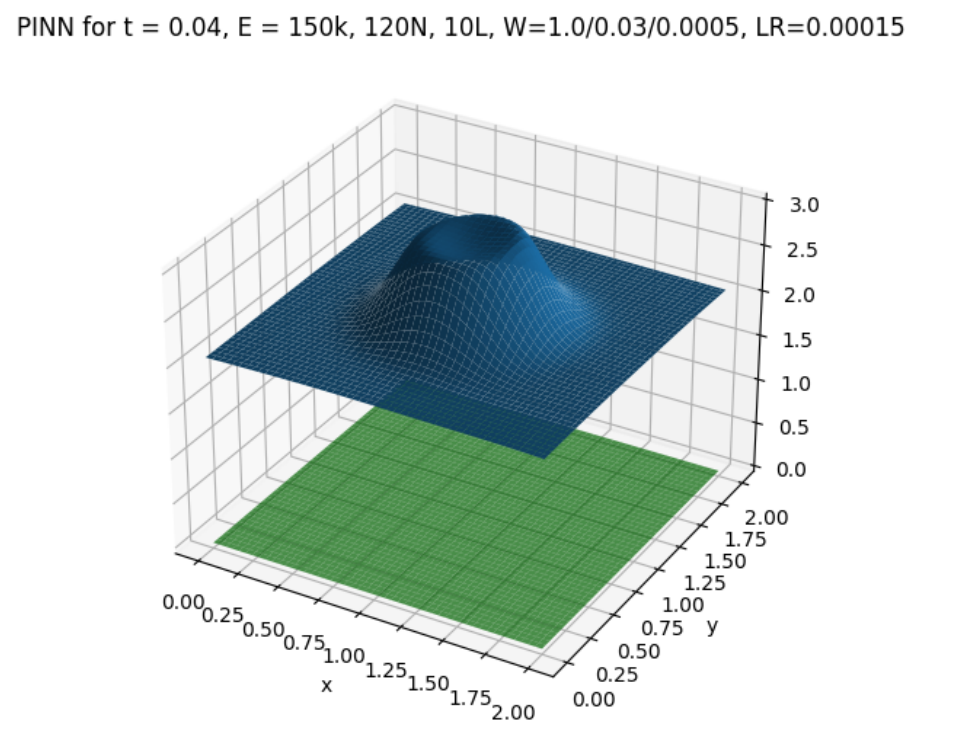} \\
      \includegraphics[width=0.4\textwidth]{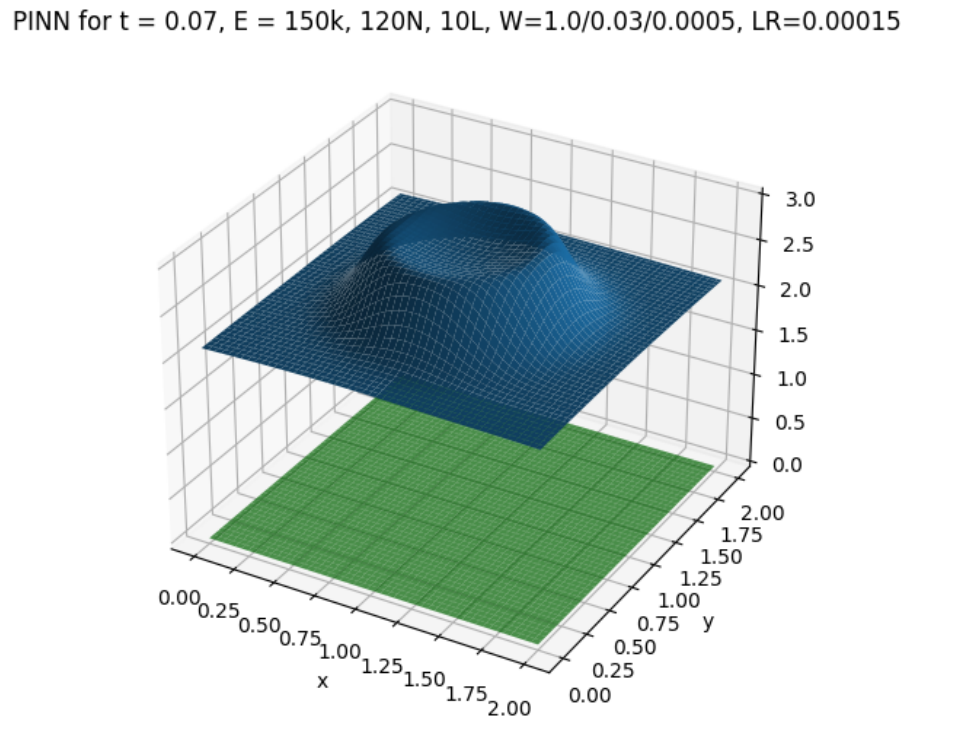} \includegraphics[width=0.4\textwidth]{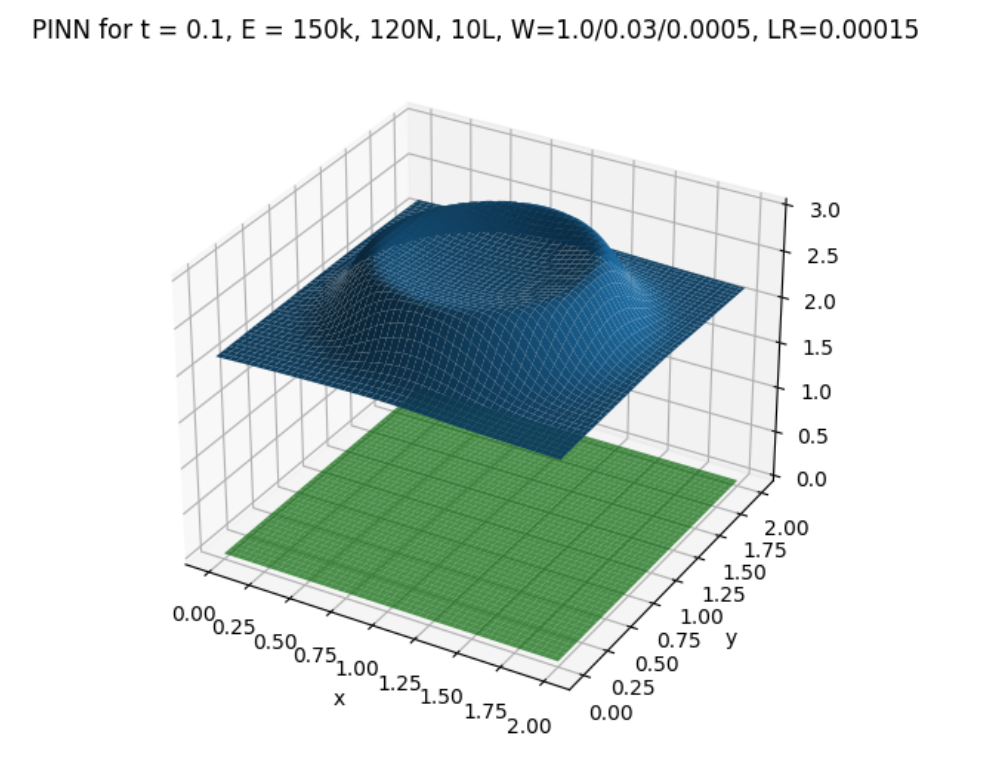} \\
      \includegraphics[width=0.4\textwidth]{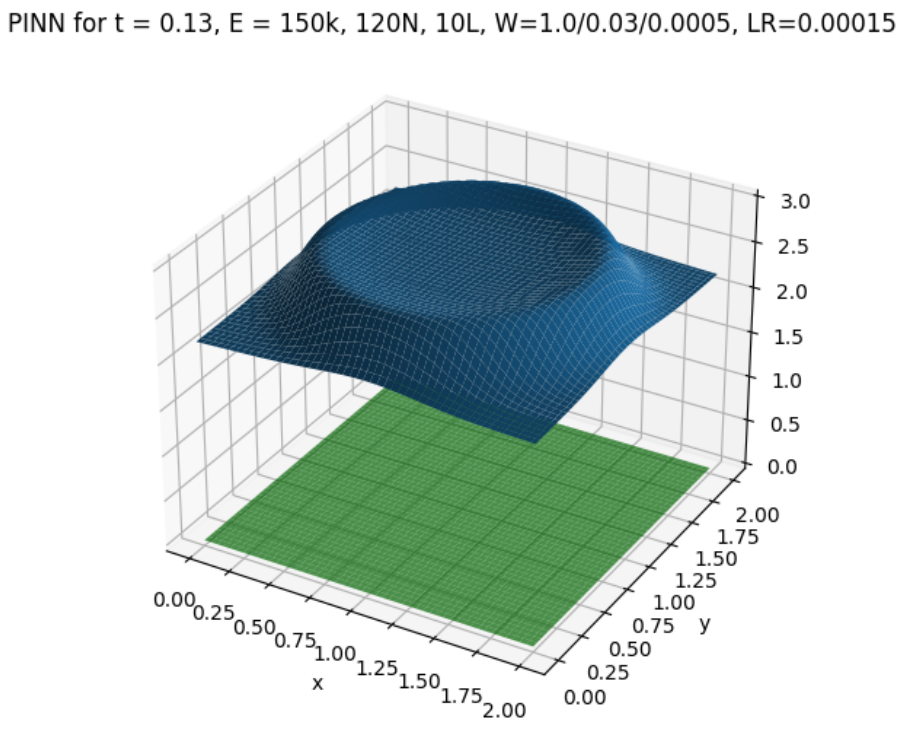} \includegraphics[width=0.4\textwidth]{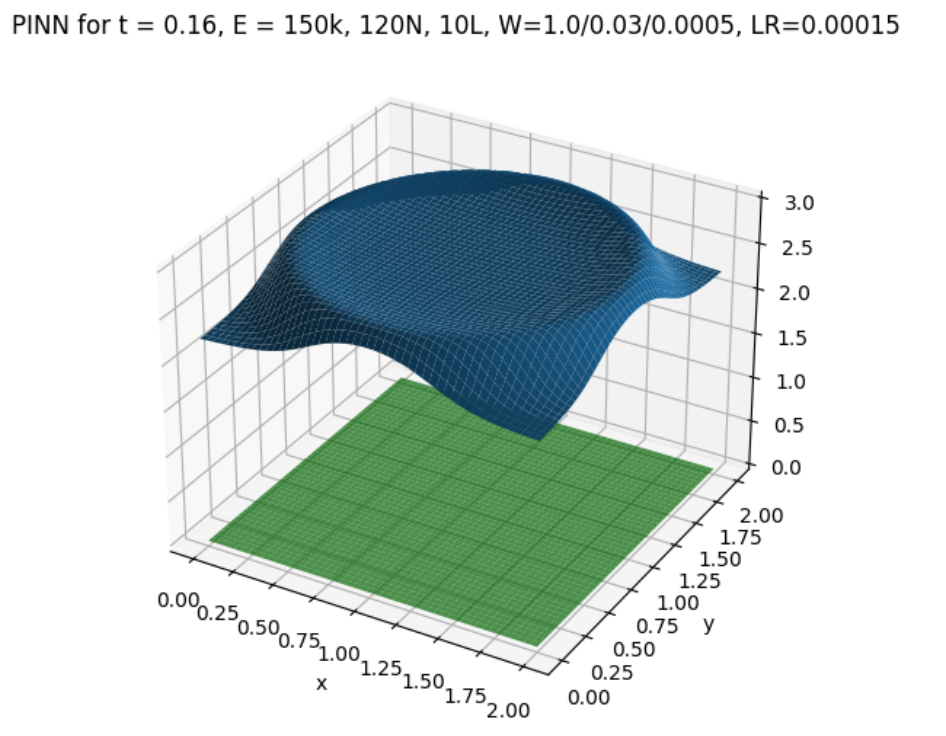} \\
      \includegraphics[width=0.4\textwidth]{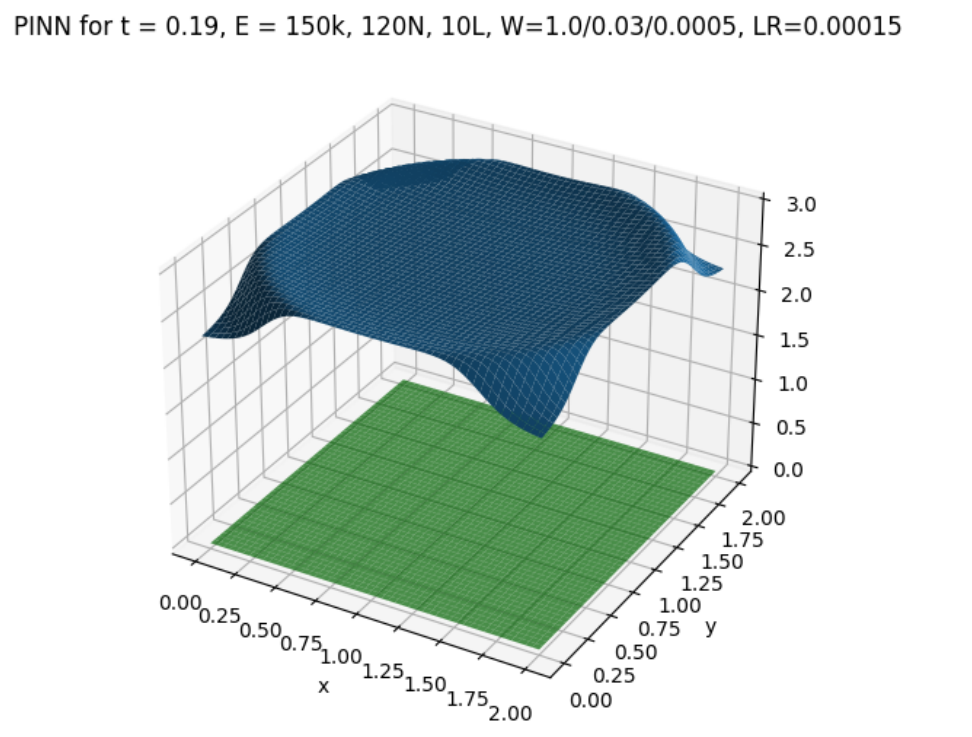} \includegraphics[width=0.4\textwidth]{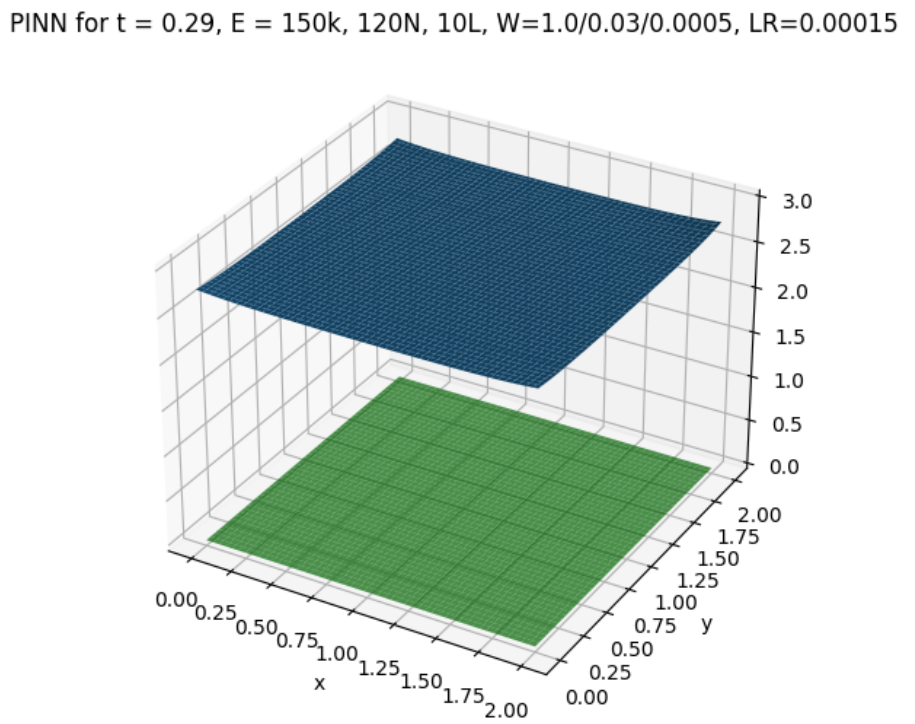}
      \caption{Wave equation simulation.}
      \label{fig:wave}
\end{figure}

The convergence of the loss is summarized in Fig. \ref{fig:losses}. The snapshots of the simulation are presented in Fig. \ref{fig:wave}.

\subsection{Thermal inversion}

In this example, we aim to model the thermal inversion effect. The numerical results presented in this section are the PINN version of the thermal inversion simulation performed using isogeometric finite element method code \cite{IGAADS} described in \cite{ShockWave}.
The scalar field $u$ in our simulation represents the water vapor forming a cloud.
The source represents the evaporation of the cloud
evaporation of water particles near the ground.
The thermal inversion effect is obtained by introducing the advection field as the gradient of the temperature.
Following \cite{Atmosphere} we define $b(x,y,t)=\frac{\partial T}{\partial y}=-2$ for lower half of the domain ($y<0.5$), and $b(x,y,t)=\frac{\partial T}{\partial y}=2$ for upper half of the domain ($y>0.5$). This represents the thermal inversion effect. The low temperature air mass is kept close to the ground by the higher temperature air mass above it. The temperature gradient decreases from the ground level to half of the domain, representing the low temperature air mass, and then the temperature gradient increases, which represents the temperature increase in the higher temperature air mass. The horizontal diffusion coefficient $K_x=0.1$ is stronger than the vertical diffusion coefficient $K_y=0.01$. The source function $f(x,y,t)$ represents the cloud vapor that evaporates from the ground level. 

We focus on advection-diffusion equations in the strong form. We seek the cloud vapor concentration field $[0,1]^2\times [0,1] \ni (x,y,t) \rightarrow u(x,y,t) \in {\cal R}$  
\begin{eqnarray}
\frac{\partial u(x,y,t)}{\partial t}+  \left( b(x,y,t) \cdot \nabla \right)u(x,y,t)- \nabla \cdot \left(K \nabla u(x,y,t)\right)  = f(x,y,t),
 \textrm{ } (x,y,t) \in \Omega \times (0,T] \\
\frac{\partial u(x,y,t)}{\partial n} = 0, \textrm{  } (x,y,t) \in \partial \Omega \times (0,T] \\
u(x,y,0) = u_0(x,y), \textrm{  } (x,y,t) \in \Omega \times 0 \\
\end{eqnarray}
This PDE translates into
\begin{eqnarray}
\frac{\partial u(x,y,t)}{\partial t}+  \frac{\partial T(y)}{\partial y}\frac{\partial u(x,y,t)}{\partial y} - 0.1\frac{\partial  u(x,y,t)}{\partial x^2}-0.01\frac{\partial  u(x,y,t)}{\partial y^2}  = f(x,y,t),
 \textrm{ } (x,y,t) \in \Omega \times (0,T] \\
\frac{\partial u(x,y,t)}{\partial n} = 0, \textrm{  } (x,y,t) \in \partial \Omega \times (0,T] \\
u(x,y,0) = u_0(x,y). \textrm{  } (x,y,t) \in \Omega \times 0 \\
\end{eqnarray}
In PINN, the neural network represents the solution,  
\begin{eqnarray}
u(x,y,t)=PINN(x,y,t)=A_n \sigma\left(A_{n-1}\sigma(...\sigma(A_1[x,y,t]+B_1)...+B_{n-1}\right)+B_n
\end{eqnarray}
where $A_i$ are matrices representing DNN layers, $B_i$ represent bias vectors, and $\sigma$ is the sigmoid activation function.
We define the loss function as the residual of the PDE

\begin{eqnarray}
LOSS_{PDE}(x,y,t) = \notag \\
\left(
\frac{\partial PINN(x,y,t)}{\partial t}+  \frac{\partial T(y)}{\partial y}\frac{\partial PINN(x,y,t)}{\partial y} - 0.1\frac{\partial  PINN(x,y,t)}{\partial x^2}-0.01\frac{\partial  PINN(x,y,t)}{\partial y^2}  - f(x,y,t)
 \right)^2
\end{eqnarray}

This residual translates to the following code

\begin{lstlisting}
def residual_loss(self, pinn: PINN):
     x, y, t = get_interior_points(self.x_domain, self.y_domain, self.t_domain, 
     self.n_points, pinn.device())
     loss = dfdt(pinn, x, y, t).to(device) 
     - self.dTy(y, t)*dfdy(pinn, x, y, t).to(device) 
     - self.Kx*dfdx(pinn, x, y, t,order=2).to(device)  
     - self.Ky*dfdy(pinn, x, y, t, order=2).to(device) 
     - self.source(y,t).to(device)
     return loss.pow(2).mean
\end{lstlisting}

We add the definitions of the {\tt Kx} and {\tt Ky} variables into the {\tt Loss} class. 
We do not change the implementation of the initial and boundary conditions, but we provide the definition of the initial state and forcing

\begin{lstlisting}
    def source(self,y,t):
      d=0.7
      res = torch.clamp((torch.cos(t*math.pi) - d)*1/(1-d), min = 0)
      res2 = (150 - 1200 * y) * res
      res3 = torch.where(t <= 0.3, res2, 0)
      res4 = torch.where(y <= 0.125, res3, 0)
      return res4.to(device)
\end{lstlisting}

During the training, we use the following global parameters

\begin{lstlisting}
LENGTH = 1.
TOTAL_TIME = 1.
N_POINTS = 15
N_POINTS_PLOT = 150
WEIGHT_RESIDUAL = 20.0
WEIGHT_INITIAL = 1.0
WEIGHT_BOUNDARY = 10.0
LAYERS = 2
NEURONS_PER_LAYER = 600
EPOCHS = 30_000
LEARNING_RATE = 0.002
\end{lstlisting}

\begin{figure}
      \centering
      \includegraphics[width=0.5\textwidth]{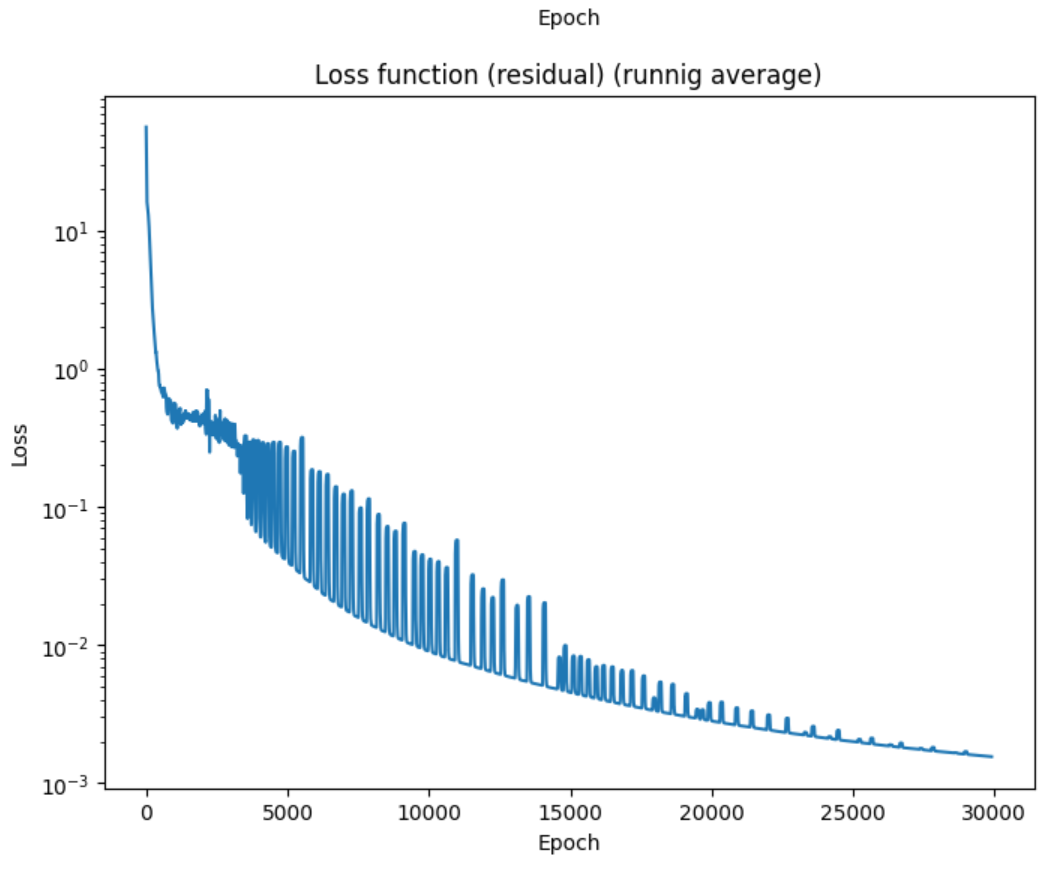}
      \caption{Thermal inversion. Convergence of the loss function.}
      \label{fig:thermal:losses}
\end{figure}

\begin{figure}
      \centering
      \includegraphics[width=0.4\textwidth]{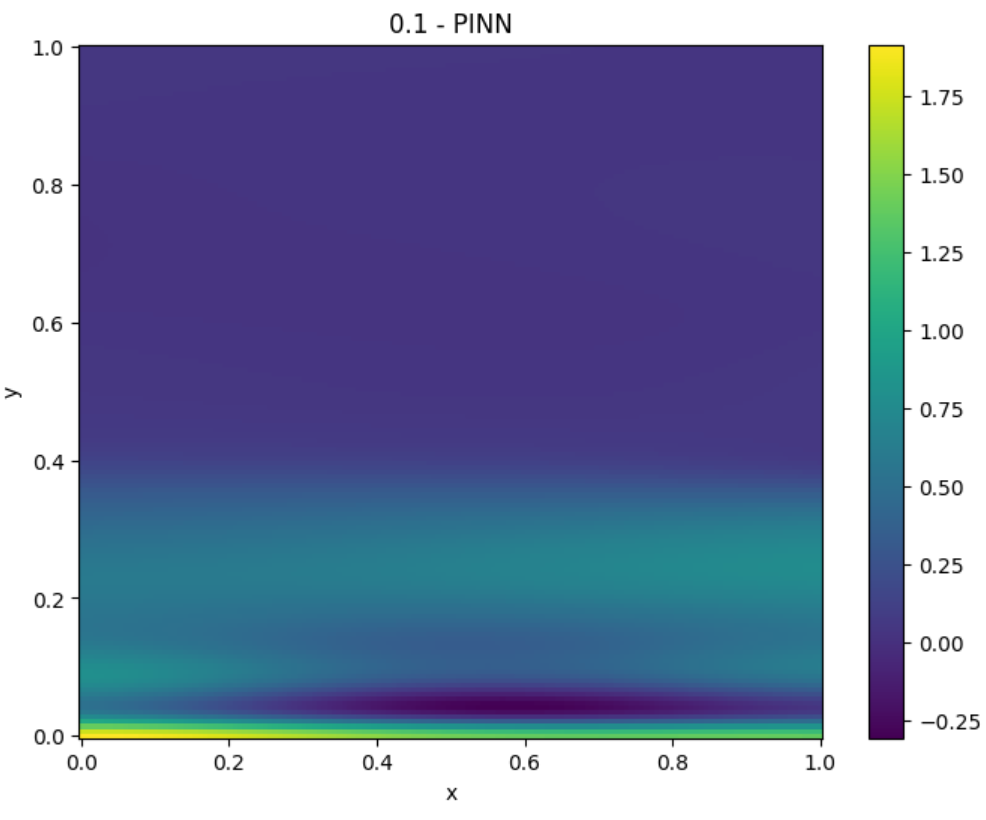}    \includegraphics[width=0.4\textwidth]{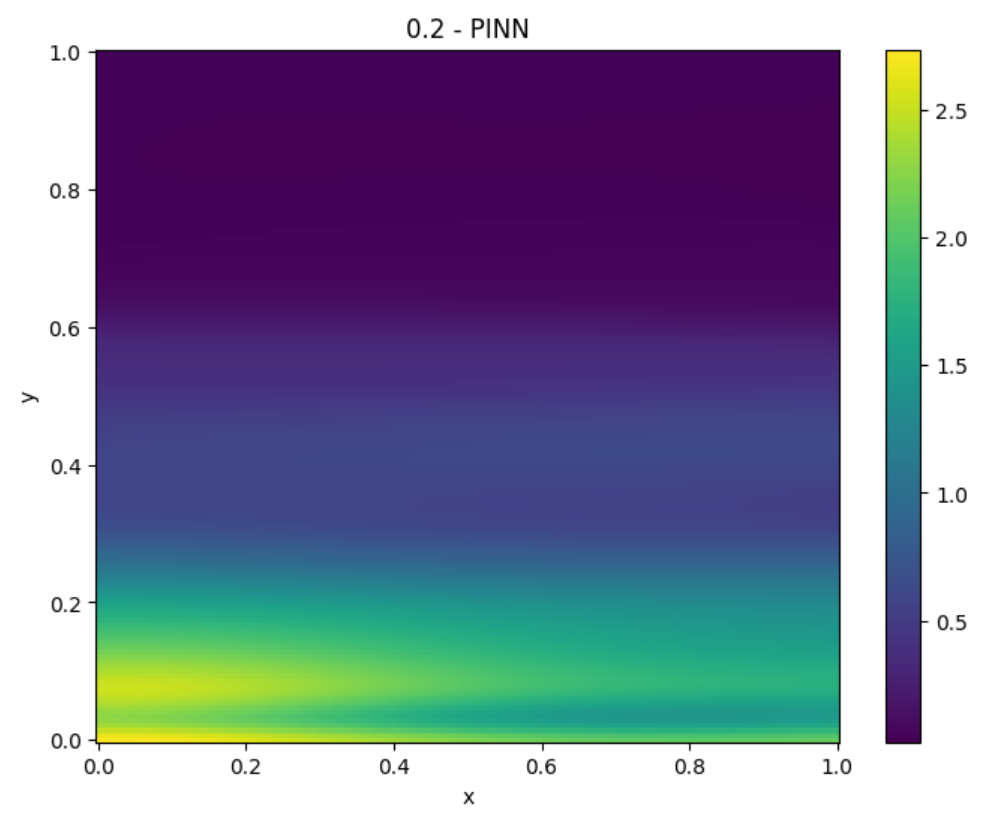} \\
      \includegraphics[width=0.4\textwidth]{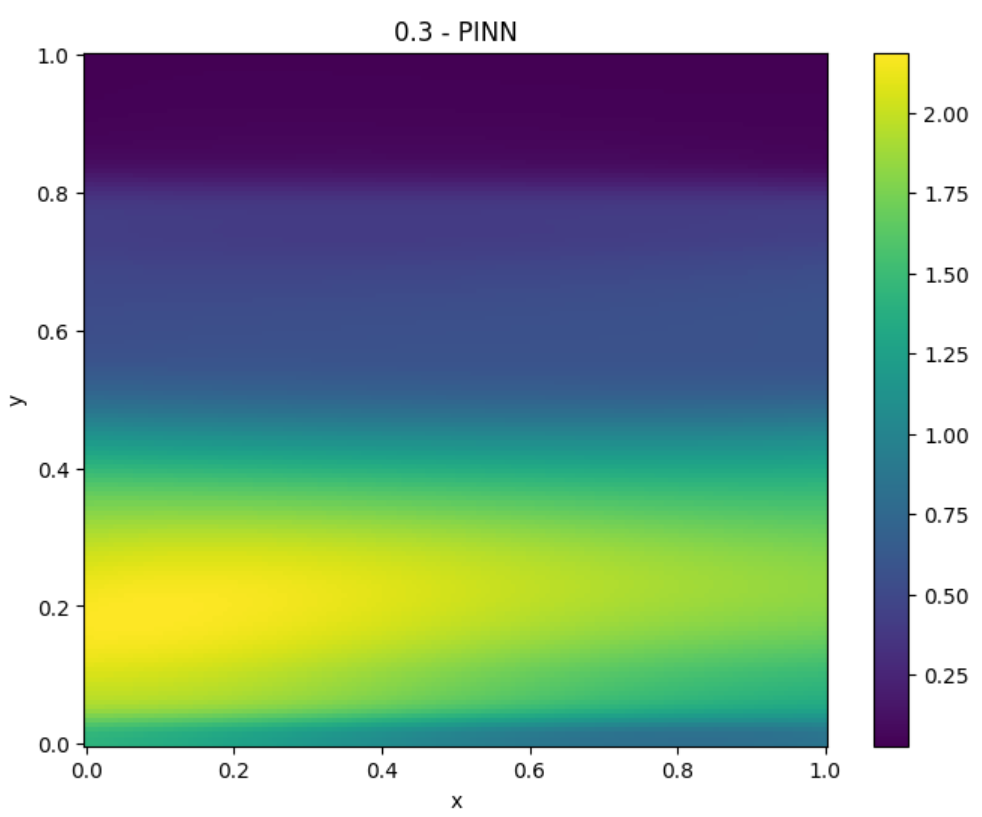} \includegraphics[width=0.4\textwidth]{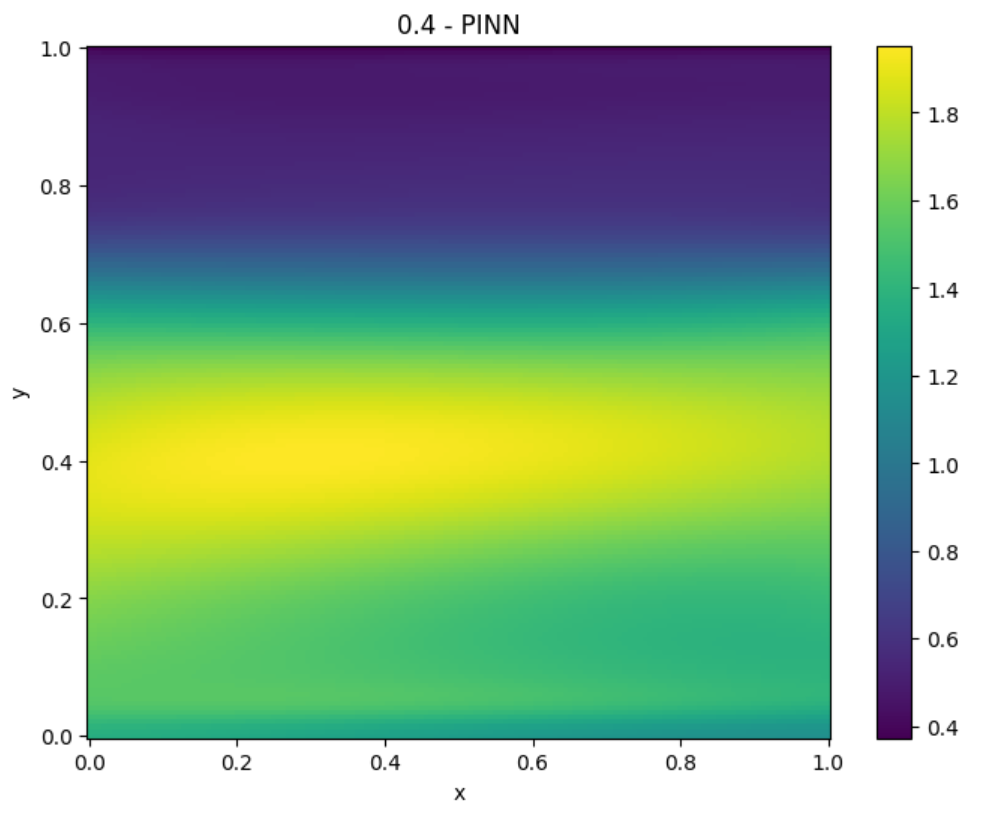} \\
      \includegraphics[width=0.4\textwidth]{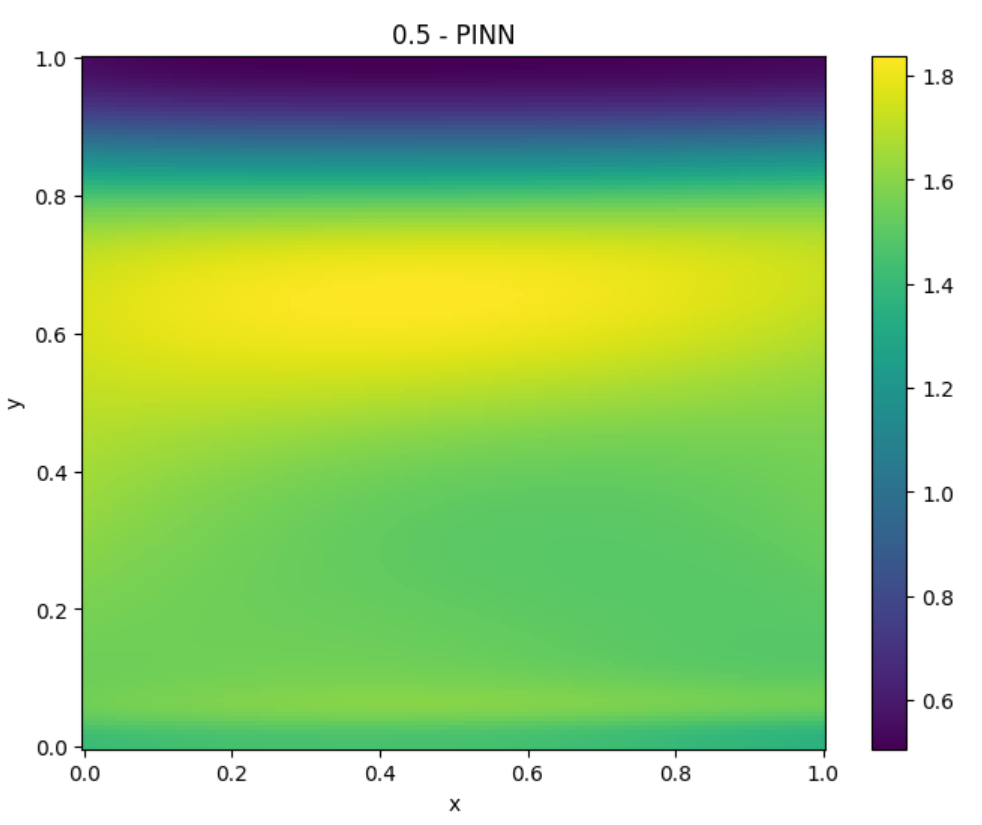} \includegraphics[width=0.4\textwidth]{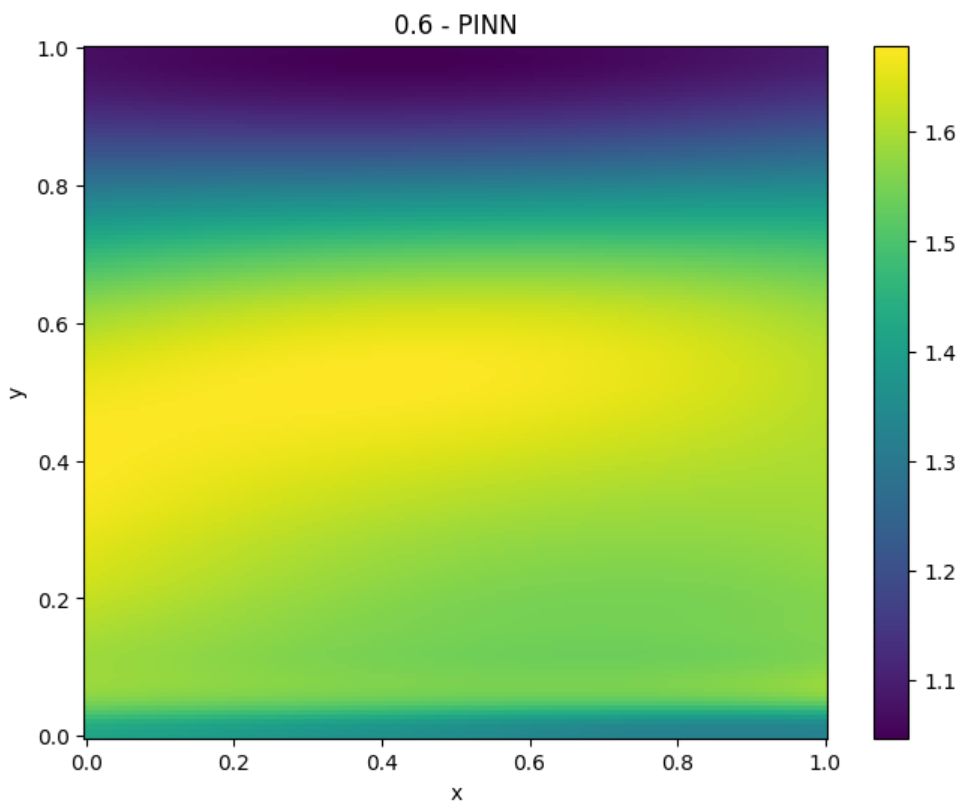} \\
      \includegraphics[width=0.4\textwidth]{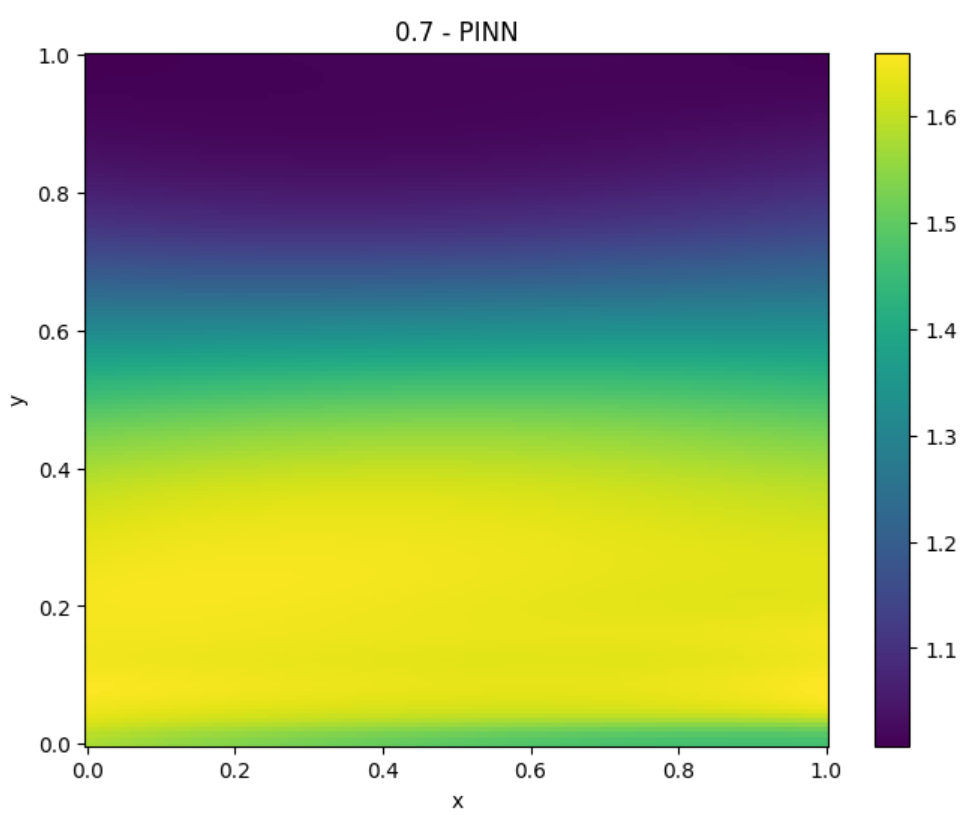} \includegraphics[width=0.4\textwidth]{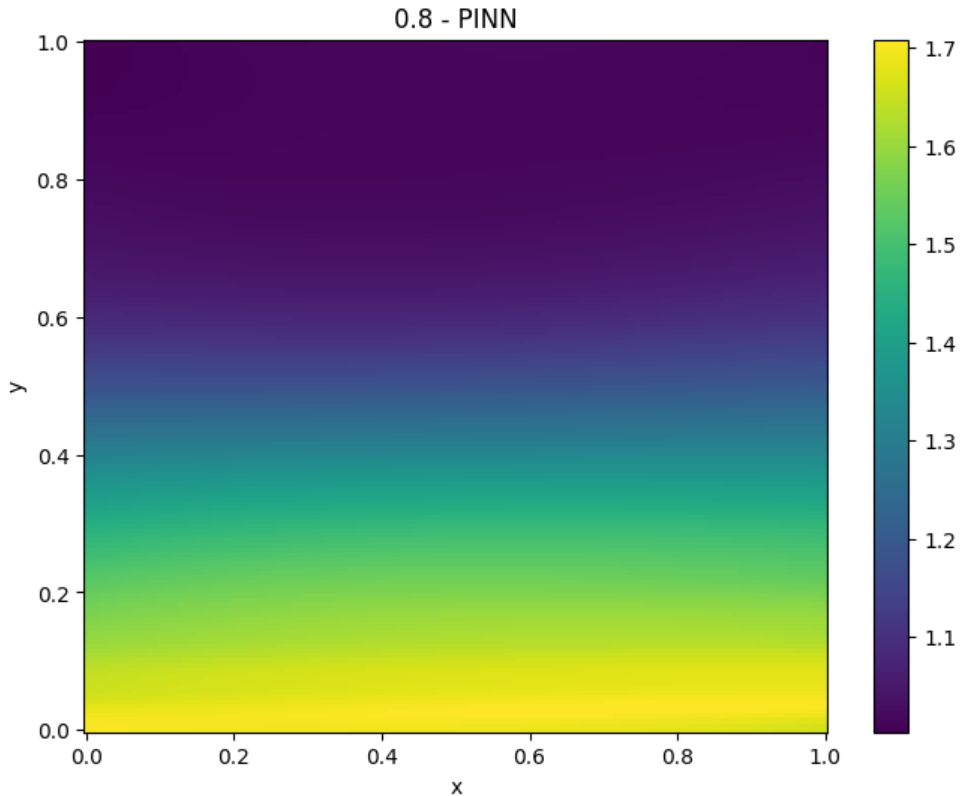}
      \caption{Thermal inversion simulation.}
      \label{fig:thermal}
\end{figure}

The convergence of the loss function is summarized in Fig. \ref{fig:thermal:losses}. The snapshots from the simulations are presented in Fig. \ref{fig:thermal}.
In the thermal inversion, the cloud vapor that evaporated from the ground stays close to the ground, 
due to the distribution of the temperature gradients.

\subsection{Tumor growth}

The last example concerns the brain tumor growth model, as described in \cite{Tumor}.
We seek the tumor cell density
$[0,1]^2\times [0,1] \ni (x,y,t) \rightarrow u(x,y,t) \in {\cal R}$,  such that

\begin{eqnarray}
\frac{\partial u(x,y,t)}{\partial t} =  \nabla  \cdot \left( D(x,y) \nabla u(x,y,t) \right) + \rho u(x,y,t) \left(1 - u(x,y,t)\right)  
 \textrm{ } (x,y,t) \in \Omega \times (0,T] \\
\frac{\partial u(x,y,t)}{\partial n} = 0, \textrm{ } (x,y,t) \in \partial \Omega \times (0,T] \\
u(x,y,0) = u_0(x,y), (x,y,t) \in \Omega \times 0 \\
\end{eqnarray}
which translates into
\begin{eqnarray}
\frac{\partial u(x,y,t)}{\partial t}-  \frac{\partial D(x,y)}{\partial x}\frac{\partial u(x,y,t)}{\partial x}
-
D(x,y)  \frac{\partial^2 u(x,y,t)}{\partial x^2} \notag \\
-
\frac{\partial D(x,y)}{\partial y}\frac{\partial u(x,y,t)}{\partial y}
-
D(x,y)  \frac{\partial^2 u(x,y,t)}{\partial y^2}
- \rho u(x,y,t) \left(1 - u(x,y,t)\right)=0
\end{eqnarray}
Here, $D(x,y)$ represents the tissue density coefficient, where $D(x,y)=0.13$ for the white matter, $D(x,y)=0.013$ for the gray matter, and $D(x,y)=0$ for the cerebrospinal fluid (see \cite{Tumor} for more details).
Additionally, $\rho=0.025$ denotes the proliferation rate of the tumor cells.
We simplify the model, by removing the derivatives of the tissue density coefficient (which is zero inside a given tissue):
\begin{eqnarray}
\frac{\partial u(x,y,t)}{\partial t}
-  D(x,y)  \frac{\partial^2 u(x,y,t)}{\partial x^2}
- D(x,y)  \frac{\partial^2 u(x,y,t)}{\partial x^y}
- \rho u(x,y,t) \left(1 - u(x,y,t)\right)=0.
\end{eqnarray}

As usual, in PINN, the neural network represents the solution,  
\begin{eqnarray}
u(x,y,t)=PINN(x,y,t)=A_n \sigma\left(A_{n-1}\sigma(...\sigma(A_1[x,y,t]+B_1)...+B_{n-1}\right)+B_n
\end{eqnarray}
with $A_i$, and $B_i$ representing  matrices and  bias vectors, and $\sigma$ is the sigmoid activation function.
We define the loss function as the residual of the PDE

\begin{eqnarray}
LOSS_{PDE}(x,y,t) = \notag \\
\left(
\frac{\partial u(x,y,t)}{\partial t}-  
D(x,y)  \frac{\partial^2 u(x,y,t)}{\partial x^2} -
D(x,y)  \frac{\partial^2 u(x,y,t)}{\partial x^y}
- \rho u(x,y,t) \left(1 - u(x,y,t)\right)
 \right)^2
\end{eqnarray}

This translates into the following code:

\begin{lstlisting}
 def residual_loss(self, pinn: PINN): 
     x, y, t = get_interior_points( 
        self.x_domain, self.y_domain, 
        self.t_domain, self.n_points, 
        pinn.device()) 
     rho = 0.025
     def D_fun(x, y) -> torch.Tensor: 
         res = torch.zeros(x.shape, dtype=x.dtype, device=pinn.device()) 
         dist = (x-0.5)**2 + (y-0.5)**2
         res[dist < 0.25] = 0.13
         res[dist < 0.02] = 0.013
         return res
     D = D_fun(x, y) 
     u = f(pinn, x, y, t) 
     loss = dfdt(pinn, x, y, t) \ 
             - D * dfdx(pinn, x, y, t, order=2) \ 
             - D * dfdy(pinn, x, y, t, order=2) \ 
             - rho * u * (1 - u) 
     return loss.pow(2).mean() 
\end{lstlisting}

The initial and boundary condition loss functions are unchanged. 
The initial state is given as follows:

\begin{lstlisting}
def initial_condition(x: torch.Tensor, y: torch.Tensor) -> torch.Tensor: 
 d = torch.sqrt((x-0.6)**2 + (y-0.6)**2) 
 res = -d**2 - 4*d + 0.4
 res = res * (res > 0) 
 return res
\end{lstlisting}
 
We setup the following model parameters

\begin{lstlisting}
LENGTH = 1.
TOTAL_TIME = 1.
N_POINTS = 20 
N_POINTS_PLOT = 150
WEIGHT_RESIDUAL = 1.0
WEIGHT_INITIAL = 1.0
WEIGHT_BOUNDARY = 1.0
LAYERS = 4
NEURONS_PER_LAYER = 80
EPOCHS = 50_000
LEARNING_RATE = 0.005
\end{lstlisting}

\begin{figure}
      \centering
      \includegraphics[width=0.6\textwidth]{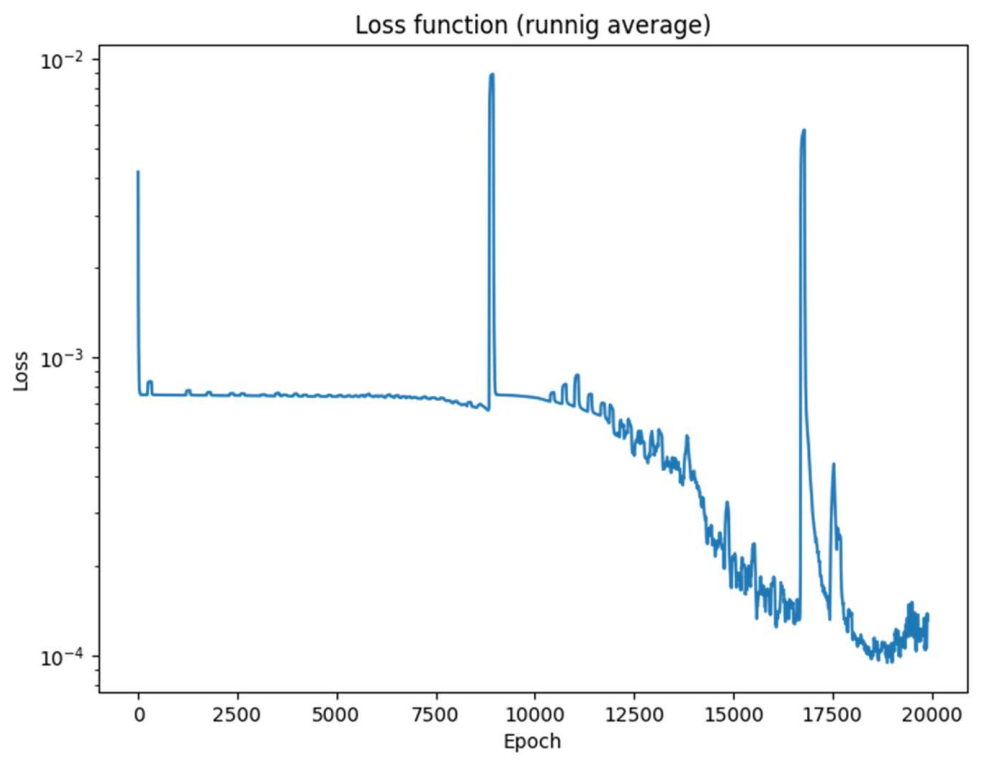}
      \caption{Tumor growth. Convergence of the loss function.}
      \label{fig:tumor:losses}
\end{figure}

\begin{figure}
      \centering
      \includegraphics[width=0.8\textwidth]{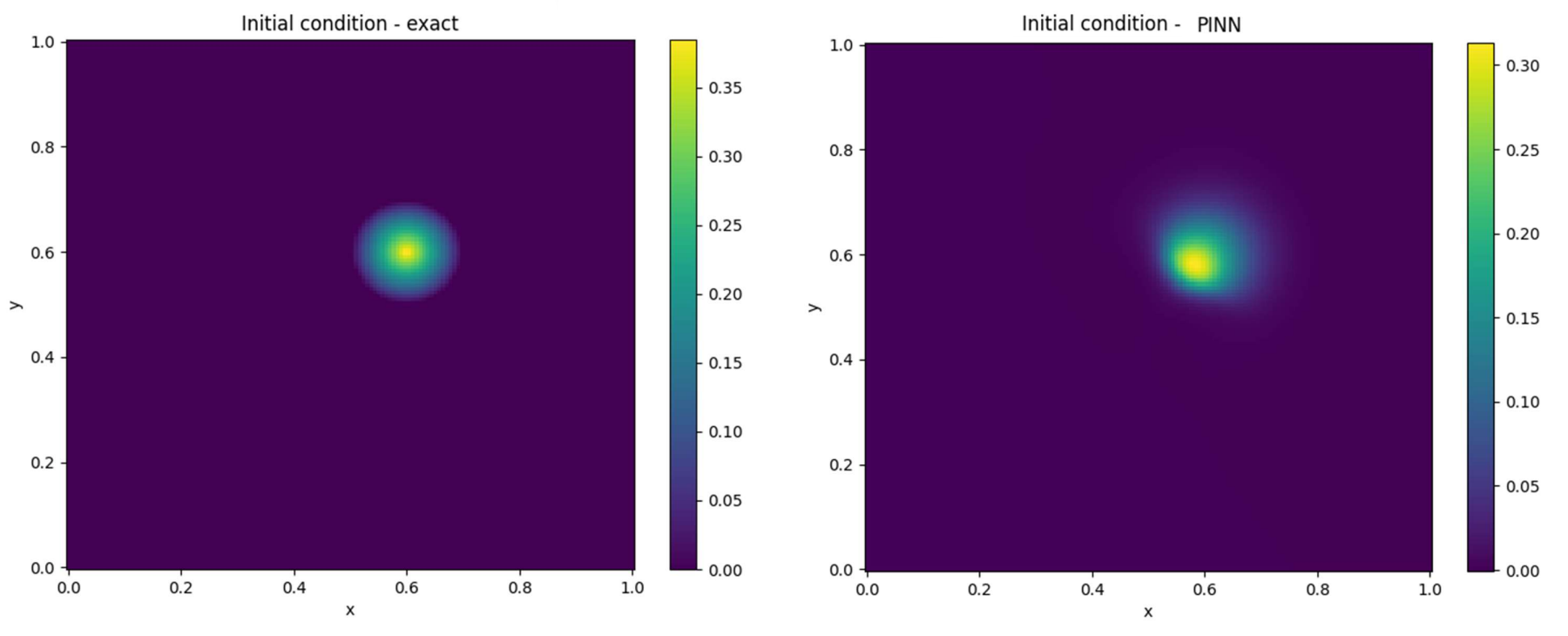}
      \caption{Tumor growth. Convergence of the loss function.}
      \label{fig:tumor:initial}
\end{figure}

\begin{figure}
      \centering
      \includegraphics[width=0.4\textwidth]{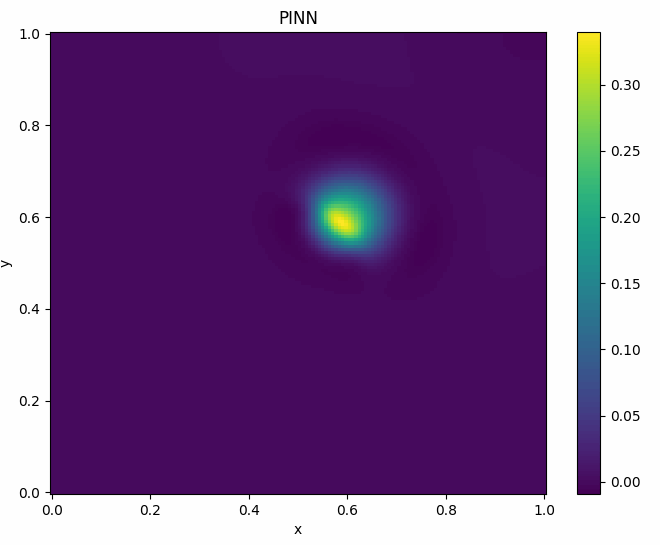}    \includegraphics[width=0.4\textwidth]{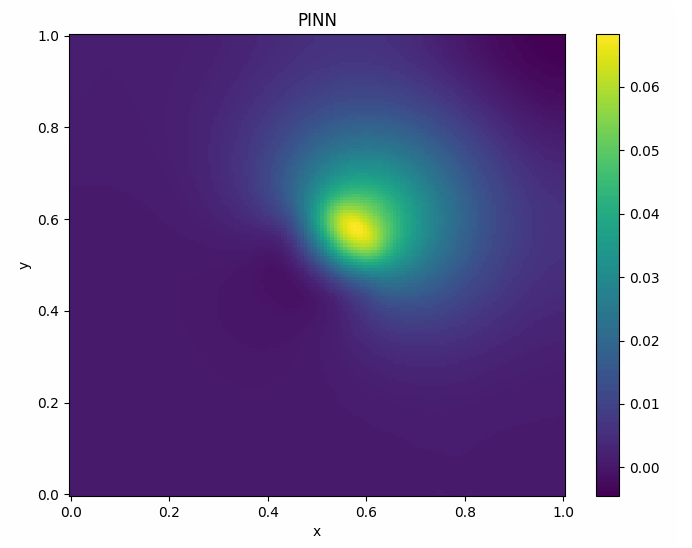} \\
      \includegraphics[width=0.4\textwidth]{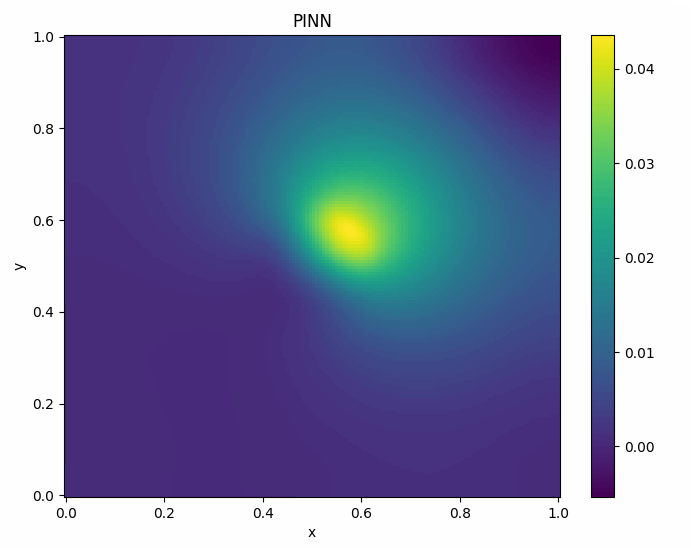} \includegraphics[width=0.4\textwidth]{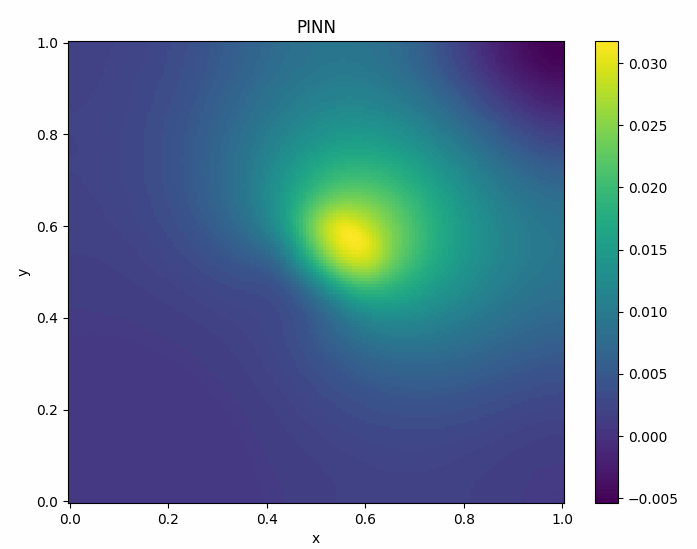} \\
      \includegraphics[width=0.4\textwidth]{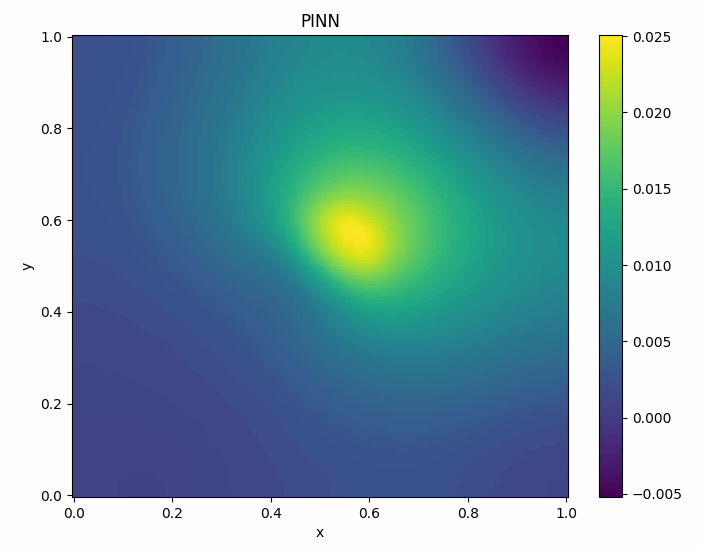} \includegraphics[width=0.4\textwidth]{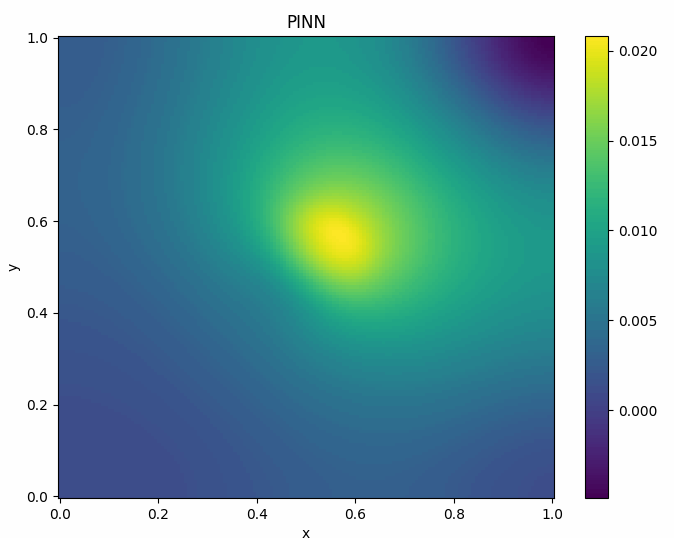} \\

      \caption{Tumor growth. Snapshots from the simulation.}
      \label{fig:tumor}
\end{figure}

We summarize in Fig. \ref{fig:tumor:losses} the convergence of the loss function. We also show how the initial data has been trained in Fig. \ref{fig:tumor:initial}. Additionally, Fig. \ref{fig:tumor} presents the snapshots from the simulation.

\clearpage

\section{Conclusions}

We have created a code
{\tt https://github.com/pmaczuga/pinn-notebooks} that can be downloaded and opened in the Google Colab. It can be automatically executed using Colab functionality. 
The code provides a simple interface for running two-dimensional time-dependent simulations on a rectangular grid.
It provides an interface to define residual loss, initial condition loss, and boundary condition loss. It provides examples of Dirichlet and Neumann boundary conditions.
The code also provides routines for plotting the convergence, generating snapshots of the simulations, verifying the initial condition, and generating the animated gifs.
We also provide four examples, the heat transfer, the wave equation, the thermal inversion from advection-diffusion equations, and the brain tumor model.

\section{Acknowledgements}

The work of Maciej Paszy\'nski, Witold Dzwinel, Pawe\l{} Maczuga, and Marcin \L{}o\'s was supported by the program ``Excellence initiative - research university" for the AGH University of Science and Technology.
The visit of Maciej Paszy\'nski at Oden Institute was partially supported by J. T. Oden Research Faculty Fellowship.

\end{document}